\documentclass[11pt,a4paper]{article}
\usepackage{jcappub}
\pdfoutput=1
\usepackage[T1]{fontenc}
\usepackage{subcaption}
\usepackage{caption}
\usepackage{tikz,xcolor,hyperref}
\usepackage{amsmath,amssymb}
\usepackage{adjustbox}
\usepackage{booktabs}
\usepackage{braket}
\usepackage{epsfig}
\usepackage{float}
\usepackage{graphicx}
\usepackage{slashed}
\usepackage[normalem]{ulem}
\usepackage{comment}
\usepackage{orcidlink}
\usepackage[svgnames]{xcolor}
\usepackage{comment}

\newcommand{\beq}{\begin{equation}}
\newcommand{\eeq}{\end{equation}}
\newcommand{\nn}{\nonumber}

\newcommand{\SD}[1]{{\color{blue}\textbf{SD:} #1}}

\begin{document}



\title{Supercooled Quality Axions}

\author[a]{Disha Bandyopadhyay\orcidlink{0009-0002-0871-8538},}
\author[a]{Debasish Borah\orcidlink{0000-0001-8375-282X},}
\author[a]{Sounak Dutta\orcidlink{0009-0007-8179-5261}}
\author[a]{and Dhruv Ringe\orcidlink{0000-0001-9052-418X}}
\affiliation[a]{\,\,Department of Physics, Indian Institute of Technology Guwahati, Assam 781039, India}
\emailAdd{b.disha@iitg.ac.in}
\emailAdd{dborah@iitg.ac.in}
\emailAdd{d.sounak@iitg.ac.in}
\emailAdd{ringedh@iitg.ac.in}

\abstract{
We study the possibility of realizing a supercooled first-order phase transition (FOPT) in a QCD axion scenario  which maintains axion quality by preventing explicit global Peccei-Quinn symmetry breaking operators upto certain order. Based on an additional $U(1)$ gauge symmetry with classical conformal invariance, we show the viability of a FOPT during radiative symmetry breaking with strong supercooling which enhances the resulting gravitational wave (GW) amplitude. A large part of the currently allowed parameter space remains within reach of current and future GW detectors like LIGO-Virgo-KAGRA (LVK), CE, ET and BBO. The small but non-zero corrections to the QCD vacuum angle also keeps complementary detection prospects promising via electric dipole moment searches in addition to generic axion searches sensitive to axion-photon coupling. A part of the parameter space also remains consistent with axion being the dark matter in the Universe.}
\maketitle

\section{Introduction}
Axion is the pseudo Nambu-Goldstone boson emerging from the spontaneous breaking of a global Peccei-Quinn (PQ) symmetry $U(1)_{\rm PQ}$ symmetry~\cite{Peccei:1977hh, Peccei:1977ur, Wilczek:1977pj, Weinberg:1977ma, Preskill:1982cy, Kawasaki:2013ae,DiLuzio:2020wdo} which was proposed to solve the strong CP problem in particle physics. The essence of the strong CP problem lies in the fine-tuning of the quantum chromodynamics (QCD) vacuum angle, \(\bar{\theta} \lesssim 10^{-10}\), required to satisfy the experimental upper bound on the neutron electric dipole moment (EDM). When the axion acquires a potential from non-perturbative QCD effects, $\bar{\theta}$ dynamically relaxes to zero, the ground state of the potential thereby solving the strong CP problem dynamically. The PQ mechanism not only provides an elegant solution to the strong CP problem, but can also solve other longstanding puzzles in particle physics and cosmology, like the origin of dark matter (DM) in the Universe \cite{Planck:2018vyg, ParticleDataGroup:2024cfk}. Depending on the specific production mechanism, ultra-light axion with sub-eV mass can constitute the entire cold dark matter (CDM) in the Universe. In the vacuum misalignment mechanism \cite{Preskill:1982cy, Abbott:1982af, Dine:1982ah}, coherent oscillations of the axion field in an expanding Universe lead the production of CDM. In order to match the observed CDM abundance, the axion decay constant is constrained to be around $f_a\sim 10^{11}$ GeV \cite{Kawasaki:2013ae}. Depending on the specific model implementation namely, Dine-Fischler-Srednicki-Zhitnitsky (DFSZ) \cite{Dine:1981rt, Zhitnitsky:1980tq} or Kim-Shifman-Vainshtein-Zakharov (KSVZ) \cite{Kim:1979if, Shifman:1979if}, additional contributions from the topological defects can alter this window by less than an order of magnitude \cite{Kawasaki:2013ae}. If the axion field has an initial non-zero velocity, as considered in the kinetic misalignment mechanism \cite{Co:2019jts, Chang:2019tvx}, the decay constant $f_a$ can be much lower while being consistent with the observed CDM relic. Axions also have promising detection prospects, with their couplings to the standard model (SM) particles, decided by the PQ symmetry breaking scale or the axion decay constant $f_a$, remain tightly constrained. For sub-eV QCD axions, there exist stringent astrophysical limits $f_a\gtrsim 10^{8\div9}$ GeV \cite{Raffelt:2006cw, Caputo:2024oqc}.

In spite of these appealing theoretical and experimental prospects of axions, the validity of the PQ mechanism is challenged by the presence of higher dimensional explicit global symmetry-breaking operators~\cite{Kamionkowski:1992mf,Holman:1992us,Barr:1992qq,Ghigna:1992iv}, particularly those induced by quantum gravity~\cite{Kallosh:1995hi, Witten:2017hdv} which is expected to violate all global symmetries explicitly. These higher dimensional operators shift the minima of the axion potential from $\bar{\theta} \simeq 0$ jeopardizing the PQ solution to the strong CP problem. For example, a dimension-five Planck-scale suppressed operator $\lambda \Phi^5 / M_{\mathrm{Pl}}$ with $f_a \sim 10^{11}~\mathrm{GeV}$ can compete with the QCD axion potential unless $\lambda \lesssim 10^{-40}$. This requires much more fine-tuning than the QCD vacuum angle in the absence of the PQ mechanism. The need for such extreme fine-tuning is often referred to as the \textit{axion quality problem}. In the literature, two broad classes of theories exist which solve the axion quality problem. The first class of such models aims to eliminate or suppress dangerous Planck-suppressed explicit PQ breaking operators. This can be achieved either by generating a naturally small $\lambda$ via non-perturbative effects, e.g., $\lambda \sim e^{-S}$ with $S \gtrsim 200$~\cite{Abbott:1989jw,Coleman:1989zu,Kallosh:1995hi,Alvey:2020nyh}, or (b) enforcing gauge symmetries that forbid operators of the form $\Delta V \sim \lambda_n \Phi^n / M_{\mathrm{Pl}}^{n-4}$ up to large $n \gg 5$ with $\lambda_n \sim \mathcal{O}(1)$~\cite{Berezhiani:1989fp, Barr:1992qq,Babu:2002ic,Qiu:2023los,DiLuzio:2017tjx,Duerr:2017amf,Fukuda:2017ylt,Ibe:2018hir,Ardu:2020qmo,Babu:2024udi,Babu:2024qzb}. Another promising way to suppress the explicit PQ-breaking contribution is through composite axion models~\cite{Randall:1992ut,Lillard:2018fdt,Gaillard:2018xgk,Vecchi:2021shj,Lee:2018yak,Cox:2023dou,Cox:2021lii,Nakai:2021nyf,Contino:2021ayn,Podo:2022gyj}. The second class of such quality axion models introduces a new QCD-like sector with a higher confinement scale $\Lambda_{\mathrm{QCD}'}$, resolving the strong CP problem in both sectors simultaneously. These models typically predict heavy, cosmologically unstable axions~\cite{Holdom:1982ex,Treiman:1978ge,Flynn:1987rs,Gherghetta:2020keg,Berezhiani:2000gh,Dimopoulos:2016lvn,Hook:2019qoh}, and evade the usual astrophysical bounds, allowing $f_a \ll 10^8~\mathrm{GeV}$, though at the cost of losing its connection to DM in the Universe. Similar ideas include multiple SM copies~\cite{Hook:2018jle,Banerjee:2022wzk}, enlarged QCD symmetries~\cite{Gherghetta:2016fhp}, and extra dimensions~\cite{Choi:2003wr,Reece:2025thc,Craig:2024dnl,Svrcek:2006yi}.

Due to the richer field content and dynamics, quality axion models often provide promising detection avenues \cite{Graham:2015ouw,FASER:2018eoc,Gligorov:2017nwh,Chou:2016lxi,Feng:2018pew,Chakraborty:2021wda,Bertholet:2021hjl,Kelly:2020dda}. Recently, there has been growing interest in exploring complementary detection prospects of such models via stochastic gravitational waves (GWs). In the quality axion models with additional $U(1)$ gauge symmetries, GWs prospects from topological defects were studied in \cite{Bandyopadhyay:2025oju, Babu:2026rbh, Zhou:2026amj} with \cite{Zhou:2026amj} also touching upon the possibility of first-order phase transition (FOPT). In the second class of models with a heavier QCD-like sector, the possibility of GWs from domain walls was studied in \cite{ZambujalFerreira:2021cte} whereas ref. \cite{Dunsky:2023ucb} studied the GW prospects from a first-order confinement transition in the heavy QCD-like sector. Motivated by these, in this work, we study the possibility of a FOPT with strong supercooling in quality axion models with additional $U(1)$ gauge symmetry. Unlike the scenario presented in \cite{Zhou:2026amj} where the GW spectrum generated by FOPT remains outside the range of any planned future experiments, here we report promising GW signatures at several future detectors like ET, CE, BBO, DECIGO with or even without supercooling. In fact, some part of the parameter space with strong supercooling also saturates the upper bounds on stochastic GWs from LIGO-Virgo-KAGRA (LVK). We also find interesting complementary detection prospects at future EDM searches and axion haloscope experiments. 

This paper is organized as follows. In section \ref{sec1}, we briefly discuss the quality axion model based on an additional $U(1)$ gauge symmetry. In section \ref{sec2}, we discuss the details of radiative symmetry breaking and supercooled first-order phase transition. In section \ref{sec3}, we show the detection prospects of the model at gravitational wave, EDM and axion search experiments. Finally, we conclude in section \ref{sec4}. In appendix \ref{appen1}, we give further details of the model related to axion quality, field-dependent masses and beta functions of different couplings followed by the details of GW production from supercooled FOPT in Appendix \ref{appen1b}. In appendix \ref{appen2}, we summarize the results for the model without classical conformal symmetry such that the spontaneous symmetry breaking occurs at tree level. 

\section{Quality axion model}
\label{sec1}
As mentioned above, we study the possibility of a supercooled FOPT in quality axion models with additional $U(1)$ gauge symmetry. While this can be achieved in any of the proposed models belonging to this category, we adopt a Babu–Dutta–Mohapatra (BDM) class of KSVZ-type axion models~\cite{Babu:2024qzb}, which is based on a gauged $U(1)_a$ extension of the SM. The gauge symmetry is $\mathcal{G}_{\rm{SM}}\times U(1)_a$, where $\mathcal{G}_{\rm{SM}} = SU(3)_c \times SU(2)_L \times U(1)_Y$ is the SM gauge group. The model has an accidental $U(1)_p\times U(1)_q$ symmetry, of which one combination is anomaly-free, and is identified as the gauge symmetry $U(1)_a$, while the other $U(1)$ has a QCD anomaly and is identified with the $U(1)_{\rm{PQ}}$ responsible for solving the strong CP problem. Two complex, SM singlet scalars, $S$ and $\phi$, are assigned $U(1)_a$ charges such that the renormalizable Lagrangian contains only terms proportional to $|S|^2$ and $|\phi|^2$, and no interlocking terms involving the two scalars. 
To allow for the possibility of a supercooled phase transition, we impose classical conformal invariance on this model so that there are no bare mass terms at the tree level. The conformal scalar potential can be written as
\beq
    V(S,\phi) = \lambda_S |S|^4 + \lambda_{\phi}|\phi|^4 
        + \lambda_{\phi S}|S|^2|\phi|^2\,.
        \label{eq: conformal_potential1}
\eeq
After spontaneous symmetry breaking, the scalars acquire vacuum expectation values (VEVs)
$$\langle S\rangle \equiv\frac{f_S}{\sqrt{2}},~\langle \phi\rangle \equiv\frac{f_\phi}{\sqrt{2}}\,.$$
One linear combination of the phases of $S$ and $\phi$ is absorbed by the $U(1)_a$ gauge boson, while the orthogonal combination remains as the physical axion, associated with the accidental PQ symmetry. The corresponding axion decay constant is related to $f_S$ and $f_\phi$ by the relation
\beq
\label{eq: fa}
f_a = \frac{f_S f_\phi}{\sqrt{9n^2 f_\phi^2 + f_S^2}}\,,
\eeq
where $n$ is an positive integer governing the $U(1)_a$ charge of $S$, $q_S=2/n$. For $f_S\simeq f_{\phi}$ we have 
\beq
\label{eq: fa_approx}
f_a \approx \frac{f_S}{3n}\,.
\eeq
For QCD axion models with sub-eV axion mass, there exists a stringent astrophysical bound on $f_a$ from SN1987A :  $f_a  \geq 4 \times 10^8 $ GeV \cite{Raffelt:2006cw, Caputo:2024oqc}. An attractive feature of this class of models is that the domain-wall number can be arranged to be $N_{\rm DW}=1$ through a suitable choice of parameters, thereby avoiding the cosmological domain-wall problem \cite{Babu:2024qzb}. A post inflationary scenario with $N_{\rm DW}=1$ allows the possibility of axion production from topological defects, suitable for axion dark matter. In the window $f_a\simeq(4.6-7.2)\times10^{10}\,\mathrm{GeV}$, axions produced from topological defects can explain $100\%$ of dark matter abundance in the Universe~\cite{Kawasaki:2014sqa}. \\


The model additionally introduces four vector-like quarks $Q_{iL,R}\,,i\in\{1,2,3,4\}$, and three right handed neutrinos (RHNs) $N_{jR}\,,j\in\{1,2,3\}$ which keep the $U(1)_a$ gauge symmetry anomaly free. The gauge charges of all fields are specified in Table \ref{table: fieldcontent}.
\begin{table}[tbp]
\centering
\renewcommand{\arraystretch}{1.3}
\begin{tabular}{ccc}
\toprule
\textbf{New fermion} &
$\mathbf{SU(3)_c \times SU(2)_L \times U(1)_Y}$ &
$\mathbf{U(1)_a}$ charge $q_i$ \\
\midrule
$(Q_L)_{1,2,3,4}$ & $(3,1,2/3)$ & $(1,1,1,-3)$ \\
$(Q_R)_{1,2,3,4}$ & $(3,1,2/3)$ & $-(1,1,1,-3)$ \\
$(N_R)_{1,2,3}$   & $(1,1,0)$ & $(2,4,-6)$ \\
\midrule
\textbf{New scalar} &
$\mathbf{SU(3)_c \times SU(2)_L \times U(1)_Y}$ &
$\mathbf{U(1)_a}$ charge \\
\midrule
$\phi$ & $(1,1,0)$ & $6$ \\
$S$ & $(1,1,0)$ & $2/n$ \\
\bottomrule
\end{tabular}
\caption{Field content and gauge charge assignments.}
\label{table: fieldcontent}
\end{table}
The chiral $U(1)_a$ charges of the fermions prohibit bare mass terms in the Lagrangian. Instead, we can write the following Yukawa terms 
\begin{align}
\label{eq: L_yuk}
-\mathcal{L}_{\mathrm{Yuk}}
&=
Y_{44}^{(Q)}\,\overline{Q}_{4L}Q_{4R}\phi^*
+
\sum_{a,b=1}^{3}
Y_{ab}^{(Q)}\,\overline{Q}_{aL}Q_{bR}
\frac{S^n}{n!\,M_*^{\,n-1}}\nn\\
&+
Y_{12}^{(N)}\,N_{1R}N_{2R}\phi^*
+
\frac{Y_{33}^{(N)}}{2}\,
N_{3R}N_{3R}
\frac{\phi^2}{2!\,M_*}
+
\mathrm{h.c.}
\end{align}
Note that the fermions $Q_{aL,R},$ for $a\in\{1,2,3\}$, and $N_{3R}$ couple to the scalars via non-renormalizable operators suppressed by a new physics (NP) scale $M_*$.
After symmetry breaking, the corresponding mass matrix is 
\beq
\label{eq: M_vlq}
(M_Q)_{ab}=\frac{M_*|Y_{ab}^{(Q)}|}{n!\,2^{n/2}}\left(\frac{f_S}{M_*}\right)^n\,.
\eeq
Due to the approximate lower bound of $m_Q\gtrsim (1-1.5)\,$TeV on VLQ masses from LHC data \cite{Benbrik:2024fku}, the NP scale can not be much larger than $f_S,~f_\phi$ for $\mathcal{O}(1)$ Yukawa couplings. Conversely, if $M_*$ is fixed, the LHC constraint translates to a lower bound on the Yukawa couplings.

Even with this enlarged symmetry and field content, it is not possible to forbid explicit PQ symmetry-breaking operators to arbitrarily higher order. The lowest-dimensional operator that is $U(1)_a$ gauge-invariant but explicitly violates the global PQ symmetry, potentially induced by quantum gravitational effects, can be written as~\cite{Babu:2024qzb}:
\begin{equation}
   V_{\rm gravity} = \kappa\, e^{i \delta}\, \frac{S^{3n} \phi^\dagger}{(3n)!\, M_{\rm Pl}^{3n - 3}} + \text{h.c.}\label{vgrav}
\end{equation}
The induced shift in the effective $\bar{\theta}$ parameter due to $V_{\rm gravity}$ can be estimated as (see Appendix \ref{appen1} for details)
\begin{equation}
    \bar{\theta} \sim \frac{\kappa \,{\rm sin\delta} \, f_S^{3n}f_\phi}{(3n)! \,2^{(3n-1)/2}\,M_{\rm Pl}^{3n-3} \, m_a^2 f_a^2}.
    \label{VPQ}
\end{equation}
The neutron electric dipole moment constraint requires the total gravity-induced QCD vacuum angle to satisfy $|\bar{\theta}_{\rm eff}|\lesssim 6\times10^{-10}$. The above expression includes only the direct gravity-induced contribution, which depends on the symmetry-breaking scales $f_S$ and $f_\phi$, the integer $n$, and the gravitational coupling. Although $n$ can be any positive integer, the tree-level axion quality requirement restricts its viable values. For generic Planck-suppressed coefficients and comparable symmetry-breaking scales, $n<3$ is excluded for $f_a\gtrsim 4\times10^8\,\mathrm{GeV}$. While $n=3$ can satisfy the tree-level constraint at smaller $f_a$, the axion dark matter window, $f_a\simeq(4.6-7.2)\times10^{10}\,\mathrm{GeV}$, which includes axion production from topological defects in post-inflationary scenarios with $N_{\rm DW}=1$~\cite{Kawasaki:2014sqa}, typically requires $n\geq4$. However, additional PQ-breaking operators involving vector-like quarks can induce loop corrections to
$\bar{\theta}_{\rm eff}$~\cite{Babu:2024qzb}. In particular, the mixed one-loop contribution can impose stronger constraints, depending on the Yukawa couplings, the NP scale $M_*$, and the symmetry-breaking scales. These contributions are discussed in Appendix~\ref{appen1}.

\section{Supercooled phase transition}
\label{sec2}
In order to realize a supercooled FOPT, we impose classical conformal invariance on the Lagrangian which gets rid of bare mass terms.  Consequently, the most general tree-level potential for the fields $S$ and $\phi$, respecting the two $U(1)$ symmetries, can be written as in Eq. \eqref{eq: conformal_potential1}
\beq
    V_0(S,\phi) = \lambda_S |S|^4 + \lambda_{\phi}|\phi|^4 
        + \lambda_{\phi S}|S|^2|\phi|^2. \label{eq: conformal_potential2}
\eeq
As we are interested in a high scale PQ phase transition, we keep the SM Higgs doublet decoupled from the singlet scalar dynamics. The quartic couplings of the singlet scalars with the SM Higgs can be tuned to generate the desired electroweak scale, as is done in generic conformal $U(1)$ gauge extensions of the SM \cite{Iso:2009ss}. We write the singlet scalar fields as
\beq
    \phi = \frac{\varphi+i \eta}{\sqrt{2}}, \qquad S=\frac{s+i \xi}{\sqrt{2}}
    \label{eq: scalar_def}\,,
\eeq
where $\eta$ and $\xi$ are Goldstone bosons associated with the two accidental $U(1)$ symmetries. We set $\eta=\xi=0$ to analyze the phase transition in the $s-\varphi$ plane. Following the Gildener–Weinberg construction \cite{Gildener:1976ih}, we seek a flat direction in field space along which the tree-level potential vanishes. For two fields, it is convenient to use polar coordinates
\beq 
\label{eq: polar}
s = r \cos\Theta\,,\quad\varphi = r \sin\Theta\,,
\eeq
such that the tree-level scalar potential takes the form
\begin{align} 
    V_0 &=\frac{r^4}{4}\left(
          \lambda_S \cos^4\Theta+\lambda_\phi \sin^4\Theta
        + \lambda_{\phi S} \cos^2\Theta \sin^2\Theta\right)\,.
\end{align}
Along the flat direction, the angular minimization condition requires
\beq
\left.\frac{\partial V_0}{\partial \Theta}\right\vert_{\Theta_{\rm{min}}} = 0\,.
\eeq 
This gives
\beq
\Theta_{\rm{min}} = \tan^{-1}\left(\sqrt{\frac{2 \lambda_S-\lambda_{\phi S}}{2\lambda_\phi-\lambda_{\phi S}}}\right)\,,\qquad V_0(r,\Theta_{\rm{min}}) = \frac{r^4}{16}\left(\frac{4\lambda_S\lambda_\phi-\lambda_{\phi S}^2}{\lambda_S+\lambda_\phi-\lambda_{\phi S}}\right)\,.
\label{eq: GW_condition1}
\eeq
We define the Gildener-Weinberg scale $\tilde\mu$ as the scale at which the tree-level potential vanishes, i.e. $V_0(r,\widetilde\Theta)=0$, where we have introduced the Gildener-Weinberg angle 
\beq
\widetilde\Theta \equiv \Theta_{\rm{min}}(\mu=\tilde\mu).
\eeq
According to Eq.\eqref{eq: GW_condition1}, this occurs when the couplings satisfy: 
\beq
\label{eq: GW_condition}
\tilde{\lambda}_{\phi S}^2 = 4 \tilde{\lambda}_\phi\tilde{\lambda}_S\,,\qquad \tilde{\lambda}_{\phi S}<0 \qquad \text{where} \,\, \tilde\lambda_i\equiv \lambda_i(\mu = \tilde\mu).
\eeq
This gives us
\beq
\widetilde\Theta = \tan^{-1}\left(\sqrt{\frac{ \tilde\lambda_S+\sqrt{\tilde\lambda_\phi\tilde\lambda_S}}{\tilde\lambda_\phi+\sqrt{\tilde\lambda_\phi\tilde\lambda_S}}}\right)
\eeq
The one-loop correction to the effective potential is given by the Coleman-Weinberg (CW) potential~\cite{Coleman:1973jx}. In the $\overline{\rm MS}$ scheme, in Landau gauge, the CW potential can be written as
\beq
V_\text{CW}(r,\Theta)=\frac{1}{64\pi^2}\sum_i(-1)^{f_i}n_im_i^4\left(r,\Theta\right)\left[ \log{\left(\frac{m_i^2\left(r,\Theta\right)}{\mu^2}\right)} - c_i \right]\,.
\eeq 
Here the sum runs over scalars $\{s,\varphi,\xi,\eta\}$, RHNs $N_{1,2,3}$, VLQs $Q_{1,2,3,4}$, and the gauge boson $X$.  $f_i=0~(1)$ for bosons (fermions), $n_i$ denotes the number of degrees of freedom (dof), and the renormalization scale is denoted by $\mu=\tilde\mu$. $m_i$ denote the field-dependent masses, and are listed in Appendix \ref{app: masses}. In the $\overline{\rm MS}$ scheme, $c_i=3/2$ for scalars and fermions and $c_i=5/6$ for gauge bosons.

Along the flat direction, masses can be obtained by setting $\Theta= \widetilde\Theta$ in their respective expressions, as shown in Appendix \ref{app: masses_flat}. For scalars, there is one massive degree of freedom corresponding to the field orthogonal to the flat direction (denoted by $s_\perp$), while the other three remain massless. Among the massless dof, two correspond to the two Goldstone bosons, while one dof corresponds to the flat direction, and is called the scalon. Then $V_{\rm{CW}}$ can be written as
\beq 
\label{eq: Vcw_flat}
V_{\rm{CW}}(r) = A r^4 + B r^4 \ln\left(\frac{r^2}{\tilde\mu^2}\right) + \mathcal{O}(r^8)\,, 
\eeq
where the first two terms include the contributions of fields that acquire masses from renormalizable terms, namely the scalar $s_\perp$, the gauge boson $X$, the RHNs $N_{1,2}$, and the VLQ $Q_4$. The $\mathcal{O}(r^8)$  contributions come from fields that acquire masses from non-renormalizable terms, i.e., RHN $N_{3}$, and VLQs $Q_{1,2,3}$. These contributions are small compared to the first two terms for $r\ll\tilde\mu$. Writing the renormalizable masses as $m_i^2(r) = \kappa_i r^2$, the coefficients $A,\,B$ in Eq.\,\eqref{eq: Vcw_flat} are given by
\begin{align}
    B &= \frac{1}{64\pi^2}\left[\kappa_{s_\perp}^2 +3\kappa_X^2-4\kappa_N^2 -12\kappa_Q^2\right]\,,\nn\\
    A &= \frac{1}{64\pi^2}\left[ \kappa^2_{s_\perp}\left(\ln \kappa_{s_\perp}-\frac{3}{2}\right) + 3\kappa^2_X\left(\ln \kappa_X-\frac{5}{6}\right)\right.\nn\\
      &\left.-4\kappa^2_N\left(\ln \kappa_N-\frac{3}{2}\right)-12\kappa^2_Q\left(\ln \kappa_Q-\frac{3}{2}\right)\right]\,,
\end{align}
where the expressions for $\kappa$ are given in Appendix\,\ref{app: masses_flat}. The scalon, which is massless at tree-level, acquires a mass radiatively. Neglecting the non-renormalizable terms, the scalon mass is given by
\beq
m^2_{\rm{scalon}} \approx 8Bf^2\,,
\eeq
where $f=\langle r\rangle \sim \tilde\mu$. A stable non-zero radiative minimum requires $B>0$. This is easily satisfied if the bosonic contributions to $V_{\rm{CW}}$ are larger than the fermionic contributions, as is the case in our model. Since the potential remains very flat near the origin, it develops a barrier at finite temperatures due to thermal contributions, thus leading to a scenario suitable for FOPT, as we discuss in details below.

\begin{table}
\centering
\setlength{\tabcolsep}{6pt}
\renewcommand{\arraystretch}{1.15}
\begin{tabular}{c|cccccc}
\hline\hline
Parameter & BP1 & BP2 & BP3 & BP4 & BP5 & BP6\\
\hline

$\tilde\lambda_S$
& 0.4274 & 0.4212 & 0.4085 & 0.2980 & 0.3932 & 0.7515\\

$\tilde\lambda_\phi$
& 0.6634 & 0.6592 & 0.6850 & 0.7173 & 0.4464 & 0.6235\\

$\tilde g_X$
& 0.1785 & 0.1666 & 0.1513 & 0.1518 & 0.1304 & 0.1365\\

$\tilde y$
& 0.2217 & 0.2182 & 0.1874 & 0.1968 & 0.1733 & 0.001732\\

$\tilde\mu$ [GeV]
& $5\times10^9$ & $5\times10^9$ & $5\times10^9$
& $5\times10^9$ & $5\times10^9$ & $7\times10^{11}$\\

$f_S/\tilde\mu$
& 1.230 & 1.264 & 1.296 & 1.444 & 1.380 & 1.04328\\

$f_\phi/\tilde\mu$
& 1.114 & 1.142 & 1.150 & 1.170 & 1.340 & 1.1014\\

$f_a/\tilde\mu$
& 0.102 & 0.105 & 0.107 & 0.120 & 0.115 & 0.0867\\

$T_c/\tilde\mu$
& 0.468 & 0.460 & 0.456 & 0.458 & 0.440 & 0.46456\\

$T_n/\tilde\mu$
& 0.0682 & 0.0514 & 0.0396 & 0.0188 & 0.0064 & 0.0546\\

$T_p/\tilde\mu$
& 0.0504 & 0.0366 & 0.0274 & 0.0123 & 0.00436 & 0.03658\\

$T_f/\tilde\mu$
& 0.0458 & 0.0328 & 0.0242 & 0.0104 & 0.00314 & 0.03144\\

$v_c/T_c$
& 3.299 & 3.429 & 3.503 & 3.749 & 4.065 & 3.0137\\

$\alpha$
& 33.979 & 119.22 & 359.19 & $9.102\times10^3$
& $4.198\times10^5$ & 100\\

$\left(\beta/H\right)_{S_3}$
& 30.53 & 26.159 & 23.201 & 19.093 & 20.668 & 18.91\\

$\left(\beta/H\right)_{RH}$
& 28.934 & 24.529 & 22.484 & 14.391 & 20.085 & 18.41\\

$R_p\,[{\rm{GeV}}^{-1}]$
& 0.239 & 0.287 & 0.322 & 0.593 & 0.400 & $2.133\times 10^{-5}$ \\

$R_0\,[\times 10^{-8}\,{\rm{GeV}}^{-1}]$
& $0.809$ & $1.064$ & $1.407$ & $2.960$ & $7.886$ & $7.366\times 10^{-3}$\\

$\bar\theta$
& $\sim10^{-30}$ & $\sim10^{-29}$ & $\sim10^{-29}$ & $\sim10^{-29}$ & $\sim 10^{-29}$ & $\sim 10^{-13}$\\

\hline\hline
\end{tabular}
\caption{
Benchmark points for $n=4$,  with progressively stronger first-order phase
transitions for BP1-BP5. BP6 is chosen to satisfy the axion dark matter constraint given in Eq.\,\ref{eq: axion_dm}. The Gildener--Weinberg scale $\tilde\mu$ is shown in
physical units, while all other dimensionful quantities are expressed as ratios to the corresponding $\tilde\mu$. The EFT hierarchy is fixed
at $M_*/\tilde\mu=20$ for all benchmark points.
}
\label{tab:benchmark_points}
\end{table}

\begin{figure}
    \centering
    \includegraphics[width=0.47\linewidth]{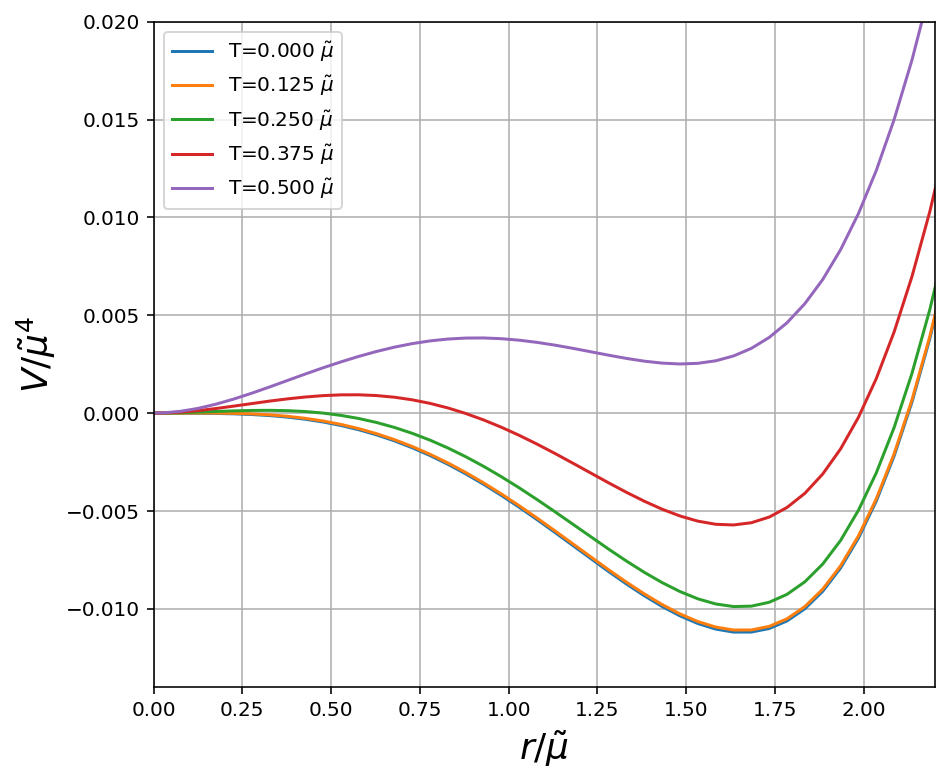}
    \includegraphics[width=0.50\linewidth]{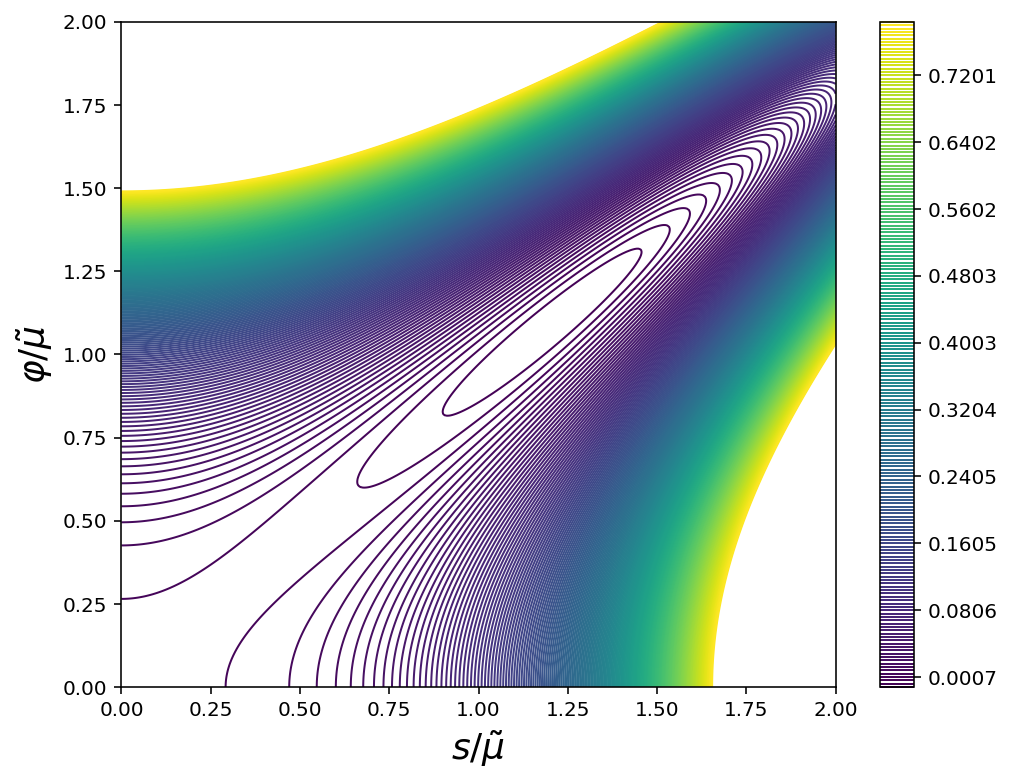}
    \caption{\label{fig: RSB} Scalar potential profile for BP1. Left panel shows the evolution of $V_{\rm{eff}}$ along the flat direction ($\Theta=\widetilde\Theta$), right panel shows the radiatively generated $T=0$ minimum in the $s-\varphi$ plane.}
\end{figure}

The above analytical expressions are useful to understand how the spontaneous symmetry breaking occurs through radiative corrections along the flat direction. However, a detailed analysis of the phase transition requires numerical computing of the full one-loop thermal effective potential in the $s-\varphi$ plane. This is given by
\beq 
V_{\rm{eff}} = V_0+V_{\rm{CW}}+V_{1T}+V_{\rm{daisy}}
\eeq
with $V_0, V_{\rm CW}$ being defined in Eq. \eqref{eq: conformal_potential2} and Eq. \eqref{eq: Vcw_flat} respectively. The one-loop thermal potential is given by~\cite{Dolan:1973qd,Quiros:1999jp}
\beq
V_\text{T}=\frac{T^4}{2\pi^2}\sum_i (-1)^{f_i} n_i J_{b/f}\left( \frac{m^2_i}{T^2} \right),
\eeq
where the definitions of $f_i, n_i$ remain same as in Eq. \eqref{eq: Vcw_flat} while $J_{b/f}$ denote the bosonic and fermionic thermal functions
\begin{equation}
    J_{b/f}(x^2) \equiv \int_{0}^{\infty} dy\, y^2 \ln\left[1 \mp e^{-\sqrt{y^2+x^2}}\right].
\end{equation}
In the Arnold-Espinosa prescription \cite{Arnold:1992rz}, the resummation of Daisy diagrams gives the Daisy contribution $V_{\rm daisy}$ to $V_{\rm eff}$, given by
\beq
\label{eq: Vdaisy}
V_{\rm{daisy}} = -\frac{T}{12\pi}\sum_i n_i \left((m_i^2+\Pi_i(T))^{3/2} - \left(m_i^2\right)^{3/2}\right)\,,
\eeq
where the sum runs over the bosonic degrees of freedom, and $\Pi_i(T)$ are the Debye masses, given in Appendix \ref{app: masses}.

Since the calculation involves multiple physical scales, we employ a renormalization-group (RG) improvement of the effective potential, with a field-dependent RG scale
\beq
\mu = {\rm{max}}(\mu_{\rm{IR}},r, T)\,,
\eeq
so that large logarithms are minimized across the relevant field and temperature ranges. The running couplings, masses, and other parameters entering $V_{\rm eff}$ are evaluated at this scale using the corresponding renormalization-group evolution (RGE) equations. The infrared scale $\mu_{\rm{IR}}$, chosen to be $10^{-6}\tilde\mu$,
prevents the RG scale from becoming arbitrarily small near the origin. We  keep the Wilson coefficients associated with the higher-dimensional fermion mass operators fixed under RGE as their anomalous dimension effects are small. The beta functions for the renormalizable couplings are given in Appendix \ref{app: beta}.

Radiative symmetry breaking leads to a non-zero minimum of $V_{\rm{eff}}(T=0)$ at
\beq 
(\langle s\rangle, \langle \varphi\rangle) \equiv (f_S,f_\phi)\,,
\eeq 
with $f_S\sim f_\phi\sim \tilde\mu$ when the self quartic couplings $\lambda_S$ and $\lambda_\phi$ are comparable. The axion decay constant is inferred from Eq.\eqref{eq: fa}. The conformal setup ensures the formation of a barrier between the origin and the non-zero minimum at finite temperatures, invariably leading to a supercooled FOPT. In some finite temperature range, true vacuum bubbles nucleate via thermal or quantum tunneling as the FOPT proceeds. For supercooled scenarios, we define four temperature scales, namely $T_c,~T_n,~T_p,~T_f$ defined as follows. At the critical temperature $T_c$ the false and true minima become degenerate. Below $T_c$, the tunneling rate per unit volume is given by
\beq
\Gamma(T) = T^4\left(\frac{S_3(T)}{2\pi T}\right)^{3/2}e^{-S_3(T)/T}\, ,
\eeq
where $S_3(T)$ is the $O(3)$ bounce action \cite{Linde:1980tt}. The nucleation temperature, $T_n$, is defined as the temperature at which the probability of nucleating one bubble per Hubble volume in unit Hubble time approaches unity, i.e.
\beq
\frac{\Gamma(T_n)}{\left(H(T_n)\right)^4}\simeq 1\, ,
\eeq
where $H(T)$ Hubble parameter at temperature $T$. The fraction of the false vacuum decreases according to 
\beq
P_f(T) = e^{-I(T)}\,,
\eeq
where the exponent is given by \cite{Guth:1979bh, Guth:1981uk}
\beq
I(T) = \frac{4\pi v_w^3}{3}\int_{T}^{T_c} \frac{dT'\,\Gamma(T')}{T'^4\,H(T')}\left[\int_{T}^{T'}\frac{d\tilde T}{H(\tilde T)}\right]^3\,.
\eeq
Here, $v_w$ is the velocity of the bubble wall. In this work, we take $v_w=1$, which is a suitable choice for strong FOPT. The percolation temperature, $T_p$, is defined by $I(T_p) = 0.34$ \cite{Ellis:2018mja}, where the false vacuum fraction reduces to $\approx 72\%$. For supercooled FOPT it is important to check that the volume of the physical false vacuum is decreasing at $T_p$, which leads to the condition
\beq
H(T_p)\left[3+ T \frac{dI}{dT}\right]<0\,.
\eeq
Finally, we define $T_f$ as the temperature at which the false vacuum fraction becomes $1\%$, indicating the end of the phase transition. The usual ordering of temperatures is $T_f<T_p<T_n<T_c$. The strength is FOPT is measured by the parameter $\alpha$, defined as
\beq 
\alpha = \frac{1}{\rho_{\rm{rad}}}\left.\left(\Delta V - \frac{T}{4}\frac{\partial\Delta V}{\partial T}\right)\right\vert_{T=T_*}\,,
\eeq
where $\rho_{\rm{rad}} = \frac{\pi}{30}g_*T^4$ is the radiation energy density of the thermal bath, $g_* \approx 160$ is the number of relativistic degrees of freedom including all the new fields beyond the SM, and $T_*$ is the temperature at which GW production is dominant. We take $T_*=T_p$ such that $\alpha \equiv \alpha_p$. 

Another crucial parameter is the rate of the phase transition given by the parameter $\frac{\beta}{H_*}$ where $H_* = H(T_*)$. For fast phase transitions, when the tunneling rate is well approximated by an exponential nucleation rate $\Gamma(t)\sim \exp(-\beta t)$, it is given by
\beq 
\frac{\beta}{H_*} = -T\frac{d}{dT}\left.\left(\frac{S_3}{T}\right)\right\vert_{T=T_*}.
\eeq
For supercooled FOPT, however, it is more appropriate to use~\cite{Megevand:2016lpr,Caprini:2024hue,Athron:2023xlk} 
\beq 
\frac{\beta}{H_*} = (8\pi)^{1/3}\frac{v_w}{R_p H(T_p)}\,,
\eeq
where $R_p$ the average bubble separation at $T_p$, given by
\beq
R_p = \frac{1}{n_b(T_p)^{1/3}}\,.
\eeq
Here $n_b(T)$ is the bubble number density at temperature $T$, given by
\beq
n_b(T) = T^3\int_{T}^{T_c}dT'\frac{\Gamma(T')P_f(T')}{T'^4H(T')}\,.
\eeq

For numerical analysis related to the FOPT, we use the \texttt{CosmoTransitions} package \cite{Wainwright:2011kj}. We first identify a few benchmark points consistent with the FOPT requirements and list them in Table \ref{tab:benchmark_points}. Fig. \ref{fig: RSB} shows the profile of the scalar potential indicating the radiative symmetry breaking for the benchmark point BP1. The left panel of the figure shows the scalar potential at different temperatures as the Universe makes a transition from the symmetric to the broken minimum. The right panel of the figure shows the contours of constant potential in the $s-\varphi$ plane clearly indicating the existence of a non-zero minima at $T=0$.

Now we perform a detailed parameter scan for points with supercooled FOPT consistent with the phenomenological constraints. First, we simplify the parameter space by taking all fermion Yukawas as universal, real, and positive, and denote them by a single parameter $y$, i.e. $$y\equiv Y^{(N)}_{12} = Y^{(N)*}_{12} = Y^{(N)}_{33} = Y^{(Q)}_{ii}\,,\quad i\in\{1,2,3,4\}\,,$$
where the vector-like quark Yukawa matrix is assumed to be diagonal. For a particular choice of the Gildener-Weinberg scale $\tilde\mu$, we use the following set of input parameters for the scan
$$\{n,\tilde\lambda_S,\tilde\lambda_\phi, \tilde g_X, \tilde y\}\,,$$
where $n\in\{3,4\}$, and the quantities with a tilde are defined at the scale $\tilde\mu$. Here $g_X$ denotes the $U(1)_a$ gauge coupling. The cross-quartic coupling $\tilde\lambda_{\phi S}$ is determined by the flatness condition given in Eq.\,\eqref{eq: GW_condition}. We also keep the NP scale $M_*$ fixed relative to $\tilde\mu$ at $M_*=20\,\tilde\mu$. For each random parameter point, we check all the phenomenological constraints. Firstly, using the astrophysical constraint $f_a\gtrsim 4\times 10^8\,{\rm{GeV}}$ mentioned earlier and using the fact that the radiatively broken minimum lies approximately at $f_S\sim f_{\phi}\sim\tilde\mu$, Eq.\,\eqref{eq: fa_approx} leads to $\tilde\mu\gtrsim 12n\times 10^8\,$GeV for a viable QCD axion model. Secondly, as the NP scale $M_*$ is held fixed relative to $\tilde\mu$, using Eq.\,\eqref{eq: M_vlq}, the LHC lower-bound of $\sim$1 TeV on the VLQ masses $m_{Q_{1,2,3}}$ translates to a lower bound on the Yukawa coupling $y$ 
\begin{align}
    y>&y_{\min}\,,\quad a\in\{1,2,3\}\,,\nn\\
    y_{\min} ={}& 2^{n/2} n!\,\left(\frac{1\,{\rm{TeV}}}{M_* }\right)\left(\frac{M_*}{f_S}\right)^n\,.
\end{align}
Finally, we also apply the neutron EDM (nEDM) bound on the QCD vacuum angle $|\overline{\theta}|\lesssim  10^{-10}$. While we do not restrict our parameter space to the one where axion abundance gives the correct DM relic in a given production mechanism, we keep in mind the preferred range
\beq
\label{eq: axion_dm}
f_a = \left(4.6-7.2\right)\times 10^{10}\,{\rm{GeV}}\,,
\eeq
where topological defects can produce the correct axion abundance. This requirement is optional whereas the first three impose stringent constraints on the parameter space. For scans without the DM requirement, we choose $\tilde\mu = 5\times 10^9\,{\rm{GeV}}$, and $M_* = 20\,\tilde\mu = 10^{11}\,{\rm{GeV}}$. For axion DM, we choose $\tilde\mu = 7\times 10^{11}\,{\rm{GeV}}$, and $M_* = 20\,\tilde\mu = 1.4\times 10^{13}\,{\rm{GeV}}$. The parameter ranges for the scans are given in Table\,\ref{tab: ranges}.  Points satisfying the above phenomenological constraints are then analyzed for supercooled FOPT and the associated phase transition parameters like $\alpha,~\beta/H_*,~T_p$ are computed.

\begin{table}[tbp]
\centering
\setlength{\tabcolsep}{6pt}
\renewcommand{\arraystretch}{1.15}
\begin{tabular}{|c|cccccc|}
\hline
 & $\tilde\mu$\,[GeV]& $n$ & $\tilde\lambda_\phi$ & $\tilde\lambda_S$ & $\tilde g_X$ & $\log_{10}\tilde y$\\
\hline
Scan1 & $5\times 10^9$ & 3 & [0,1] & [0,1] & [0,0.3] & [-4,0]\\
Scan2 & $5\times 10^9$ & 4 & [0,1] & [0,1] & [0,0.3] & [-2.5,0.5]\\
Scan3 & $7\times 10^{11}$ & 4 & [0,1] & [0,1] & [0,0.3] & [-4,0]\\
\hline
\end{tabular}
\caption{Input parameter ranges for the scans. The NP scale is set to $M_* = 20\,\tilde\mu$. For Scan3, the $\tilde\mu$ is chosen to include the axion dark matter window for $f_a$. }
\label{tab: ranges}
\end{table}


\begin{figure}
    \centering
    \includegraphics[width=0.98\linewidth]{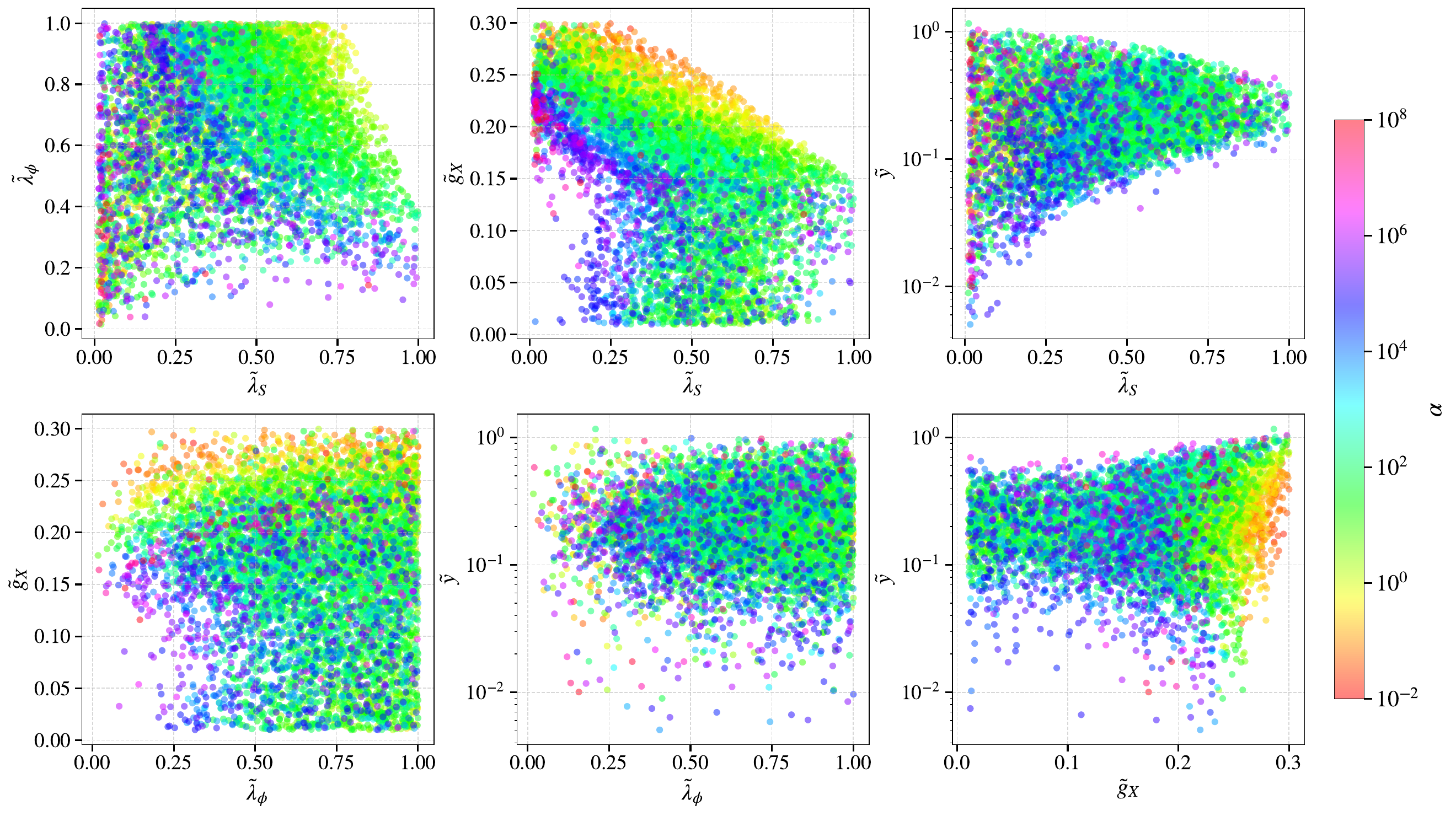}
    \caption{\label{fig: param_scan_n4}Parameter space consistent with the FOPT criteria and other phenomenological bounds for $n=4$ with the color code indicating the value of $\alpha$.}
\end{figure}


\begin{figure}
    \centering
    \includegraphics[width=0.98\linewidth]{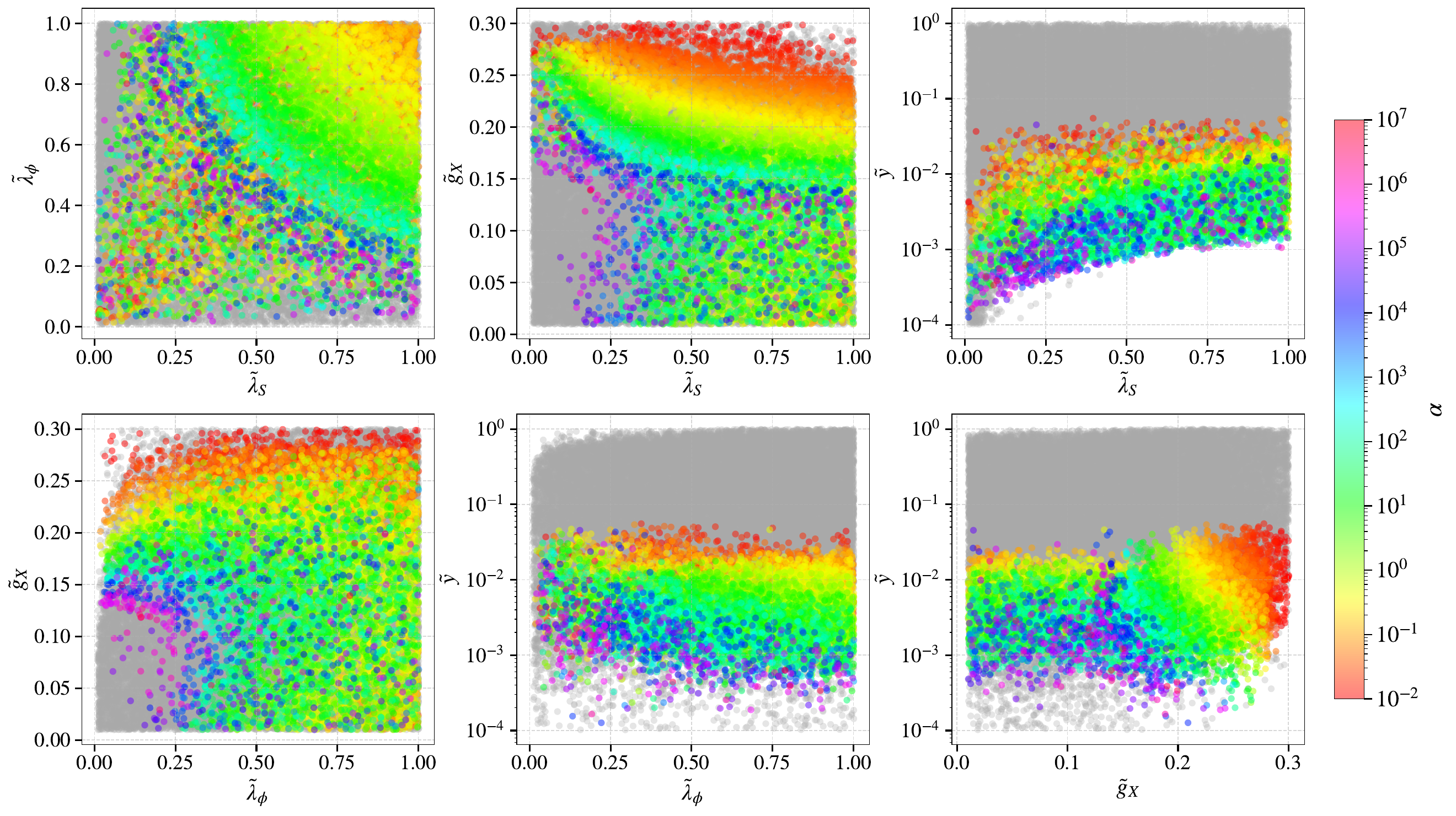}
    \caption{\label{fig: param_scan_n3} Parameter space consistent with the FOPT criteria and other phenomenological bounds for $n=3$ with the color code indicating the value of $\alpha$. The grey colored points are ruled out by current upper bounds on neutron EDM.}
\end{figure}

 \begin{figure}
    \centering
    \includegraphics[width=0.5\linewidth]{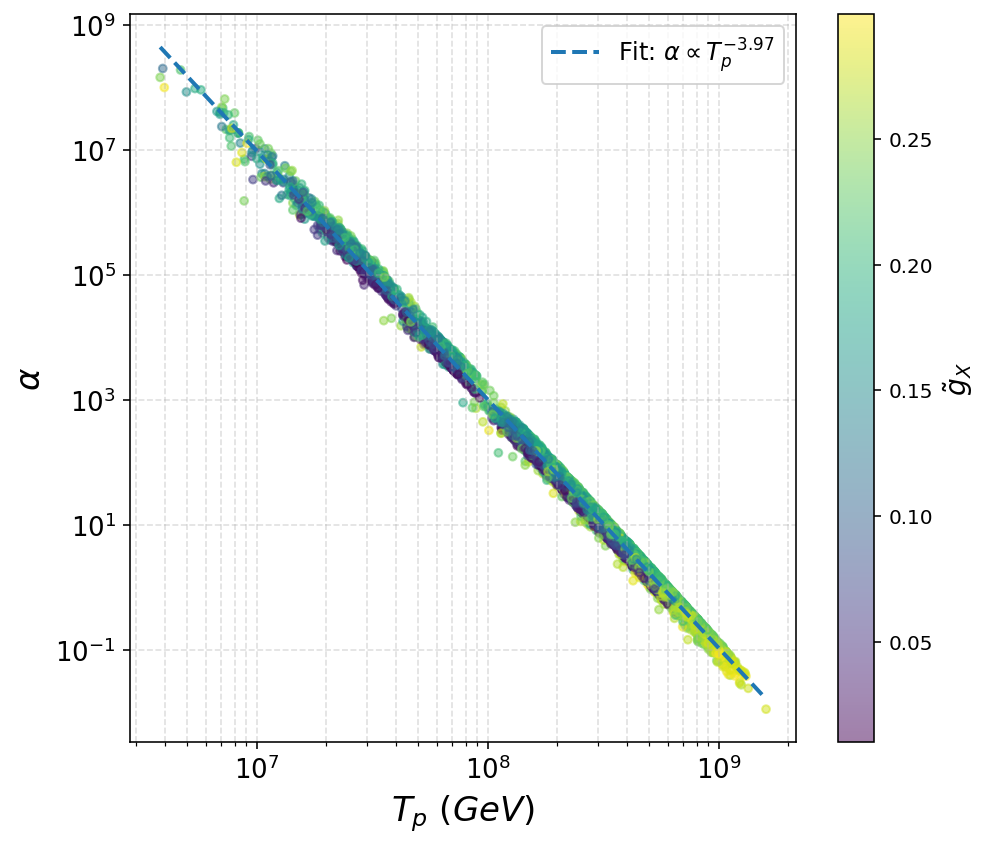}
    \caption{\label{fig: alphap_Tp} Parameter space consistent with the FOPT criteria and other phenomenological bounds for $n=3$ in $\alpha-T_p$ plane. The color code indicates the $U(1)_a$ gauge coupling strength.}
\end{figure}

\begin{figure}
    \centering
    \includegraphics[width=0.7\linewidth]{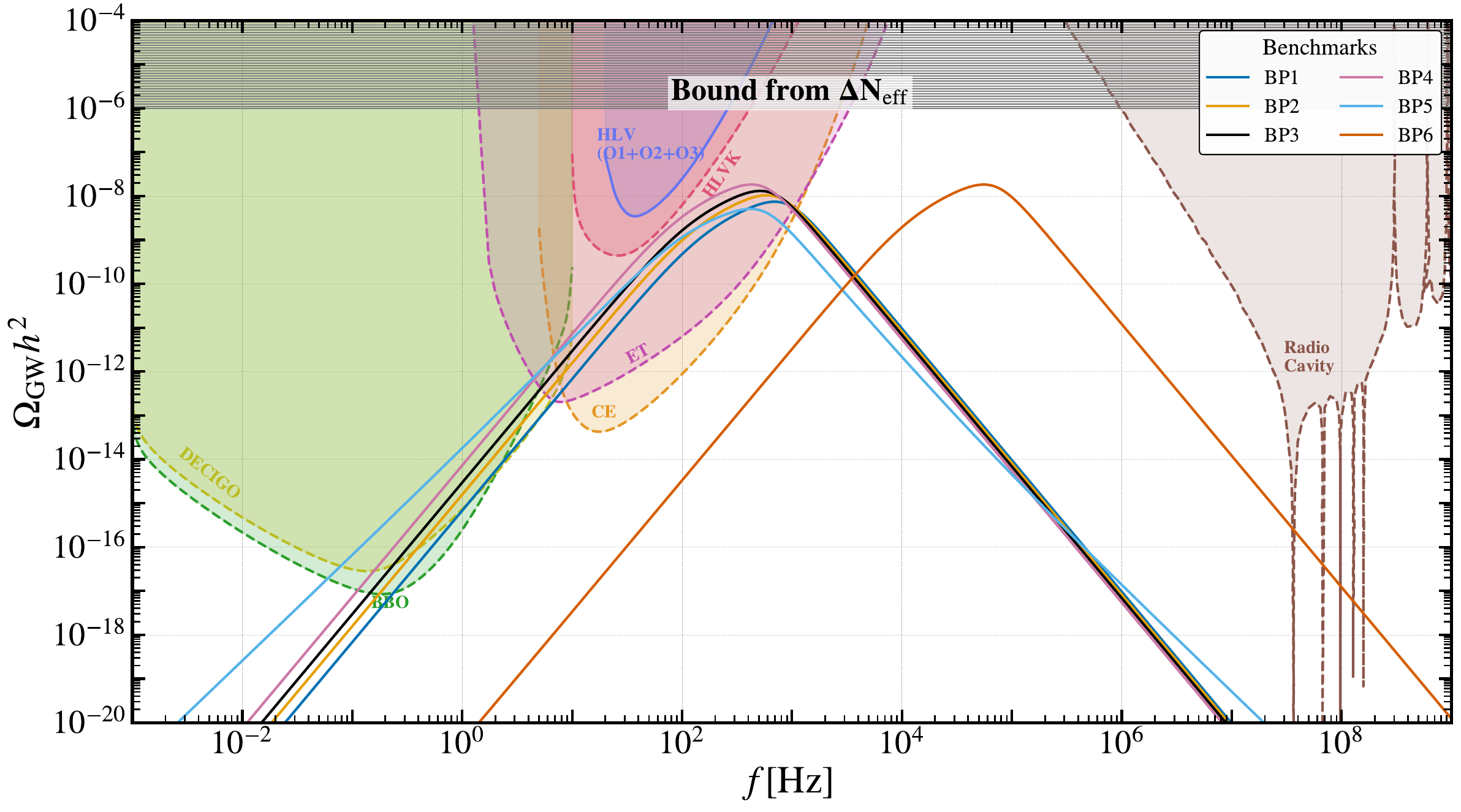}
    \caption{Stochastic gravitational wave spectrum for the benchmark points listed in Table \ref{tab:benchmark_points}. The shaded regions correspond to sensitivities of present and future GW detectors. The grey-shaded region at the top is ruled out due to upper bound on GW contribution to the effective number of relativistic degrees of freedom, $N_{\rm eff}$ \cite{Planck:2018vyg}.}
    \label{fig:GWconformalNLOSpec}
\end{figure}

Fig.\ref{fig: param_scan_n4} and Fig.\ref{fig: param_scan_n3} show the results of our numerical scan in the model parameter space for $n=4$ and $n=3$ respectively with the color code indicating the value of the FOPT parameter $\alpha$ spanning several orders of magnitude. The plots indicate a strong correlation between $g_X$ and the strength of supercooling $(\alpha)$, while $y$ is remarkably uncorrelated. Additionally, the supercooling has a weak dependence on scalar self couplings $\lambda_S, \lambda_\phi$. Large supercooling is concentrated at intermediate/small $g_X$, while large $g_X$ strongly suppresses it. Increasing $g_X$	
  substantially raises the percolation temperature, thereby reducing the amount of supercooling. Since $\alpha \propto T_p^{-4}$, this produces the dominant suppression of $\alpha$. This is also illustrated in Fig. \ref{fig: alphap_Tp}. The variation of gauge coupling, shown as color code in Fig. \ref{fig: alphap_Tp} also confirms the respective correlations observed in Fig.\ref{fig: param_scan_n4} and Fig.\ref{fig: param_scan_n3}. Interestingly, for $n=3$, a substantial part of the parameter space is already disfavored by the current nEDM bounds, indicated by the grey colored points in Fig.\ref{fig: param_scan_n3}. This brings complementary detection prospects in terms of GWs and EDM searches which we discuss in the upcoming section.

\section{Detection Prospects}
\label{sec3}
In this section, we summarize the complementary detection prospects of the model, particularly focussing on stochastic GWs, axion detection and EDM searches.

For a first order phase transition, the stochastic gravitational wave spectrum is made up of three different contributions namely, the bubble collisions~\cite{Turner:1990rc,Kosowsky:1991ua,Kosowsky:1992rz,Kosowsky:1992vn,Turner:1992tz}, the sound wave of the plasma~\cite{Hindmarsh:2013xza,Giblin:2014qia,Hindmarsh:2015qta,Hindmarsh:2017gnf} and the turbulence of the plasma~\cite{Kamionkowski:1993fg,Kosowsky:2001xp,Caprini:2006jb,Gogoberidze:2007an,Caprini:2009yp,Niksa:2018ofa}. For our study, we use the numerical fits adopted by the LISA Cosmology working group~\cite{Caprini:2024hue,Matuszak:2026xsz}. The total contribution to the stochastic GW background redshifted to today is given by
\begin{equation}
\label{eqGW1}
h^2 \Omega_{\mathrm{GW},0}
= \mathcal{R} \left(
h^2 \Omega_{\mathrm{col}}
+ h^2 \Omega_{\mathrm{sw}}
+ h^2 \Omega_{\mathrm{turb}}
\right),
\end{equation}
where $\mathcal{R}$ is the redshift factor. We also consider the next-to-leading-order (NLO) corrections to the GW spectrum for better accuracy of our results. The details of these individual contributions and NLO corrections can be found in Appendix \ref{appen1b}. In Fig.~\ref{fig:GWconformalNLOSpec}, we show the GW spectra for the benchmark points listed in Table \ref{tab:benchmark_points}. The colored contours correspond to the power law integrated (PLI) sensitivities of upcoming and proposed GW detectors, such as BBO~\cite{Corbin:2005ny,Crowder:2005nr}, DECIGO~\cite{Kawamura:2020pcg,Kudoh:2005as,Yagi:2011wg}, ET~\cite{Hild:2008ng}, CE~\cite{LIGOScientific:2016wof} and future upgrades of LVK~\cite{KAGRA:2021kbb,Jiang:2022uxp,LIGOScientific:2022sts}. We have also included the projected sensitivity of resonant electromagnetic cavities (Radio Cavity) adopted from~\cite{Herman:2022} \footnote{We note that these sensitivity projections are criticized in the literature as being overly optimistic~\cite{Aggarwal:2025noe}.}.  The PLI sensitivities for BBO, DECIGO, CE, ET and HLVK are obtained from Ref.~\cite{Schmitz:2020syl} and its accompanying numerical repository~\cite{Schmitz:Zenodo}. For the LIGO-Hanford, LIGO-Livingston, and Virgo (HLV) network, we utilize the publicly available data from the combined O1, O2, and O3 observing runs~\cite{LIGO:O3Stochastic,LIGO:HLVData}. From Fig.~\ref{fig:GWconformalNLOSpec}, we observe that only for \(f_a\) close the the astrophysical lower limit on the PQ breaking scale for QCD axion (\(\sim 4\times 10^8~{\rm GeV}\)), the GW signals for the supercooled FOPT can be probed by ET, CE, DECIGO and BBO.  We note that the points showing maximum supercooling typically have large GW peak amplitudes. A higher value of \(\alpha(T_p)\) indicates a stronger FOPT, and gives a larger GW amplitude. We also prefer lower values of \(\beta/H_p\) and \(T_p\) (and consequently, \(f_a\)) to have the points fall within the detectability region. While BP5 has higher $\alpha$, its peak amplitude remains lower due to larger \(\beta/H_p\). On the other hand, the GW spectrum for the benchmark point BP6 having higher \(f_a\) fall into a desert region where there exist no planned detectors. In order check the GW prospects of the entire scanned parameter space, we also compute the signal-to-noise ratio (SNR) for a particular experiment, the details of which can be found in Appendix \ref{appen1b}.

\begin{figure}[H]
    \centering
    \includegraphics[width=0.7\linewidth]
    {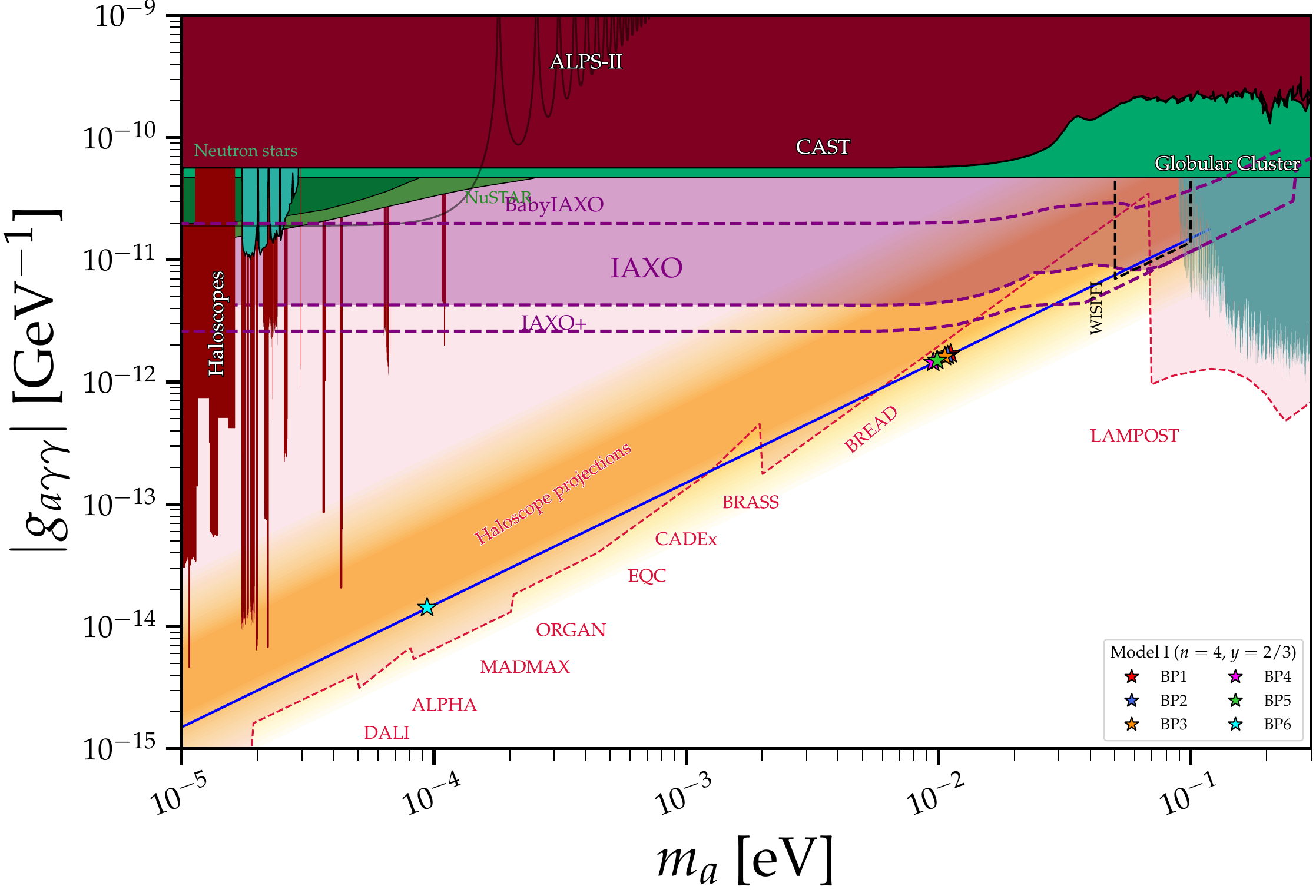}
    \caption{Existing experimental constraints and projected sensitivities in the axion mass $m_a$ versus axion--photon coupling $|g_{a\gamma\gamma}|$ plane. The solid blue line corresponds to Model I with vector-like quark hypercharge $Y=2/3$, yielding $E/N=8/3$, which coincides with the conventional DFSZ-I anomaly ratio. The differently coloured star markers denote the six benchmark points (BP1--BP6) for $n=4$, with their corresponding model parameters given in Table~\ref{tab:benchmark_points}. The experimental bounds and projected sensitivities are taken from the AxionLimits compilation ~\cite{AxionLimits}.}
    \label{fig:det}
\end{figure}

The axion detection prospects in most experiments primarily rely on axion-photon coupling. The axion-photon interaction is parametrized as 
\begin{equation}
  \mathcal{L}_{a\gamma \gamma} \supset \frac{1}{4} g_{a \gamma \gamma} \, \phi_a \,  F_{\mu \nu} \tilde{F}^{\mu \nu},
\end{equation}
where, $g_{a \gamma \gamma } = (\frac{E}{N} - 1.92)\frac{\alpha}{2 \pi f_a}$ \cite{Srednicki:1985xd} with $E, N$ representing electromagnetic and color anomaly coefficients, respectively. Fig.~\ref{fig:det} shows the parameter space in the axion--photon coupling $|g_{a\gamma \gamma}|$ versus axion mass $m_a$ plane, together with the existing experimental constraints and projected sensitivities of various axion searches. The solid blue line corresponds to the BDM model of \cite{Babu:2024qzb} with the vector-like quark hypercharge $Y=2/3$, yielding an electromagnetic-to-color anomaly ratio $E/N=8/3$. This anomaly ratio coincides with that of the conventional DFSZ-I axion model, although the underlying particle content and PQ charge assignments are different. The six benchmark points, BP1--BP6, indicated by differently colored $\star$ markers, correspond to the parameter choices listed in Table~\ref{tab:benchmark_points} for $n=4$. Their positions are determined by the axion decay constant $f_a$, calculated from the scalar vacuum expectation values $f_S$ and $f_\phi$, which fix both the axion mass and its photon coupling. While BP1--BP5 are clustered around $m_a\sim10^{-2}\,\mathrm{eV}$, BP6 lies at a significantly lower mass, $m_a\sim8\times10^{-5}\,\mathrm{eV}$, owing to its larger symmetry-breaking scale. The existing bounds and projected sensitivities shown are taken from the \texttt{AxionLimits}
compilation~\cite{AxionLimits}, which provides the corresponding
data and references to the original studies. The shaded exclusion regions enclosed within solid contours represent existing bounds, while the regions enclosed by dashed contours indicate projected sensitivities of future searches. 
\begin{figure}
    \centering
    \includegraphics[width=0.5\linewidth]{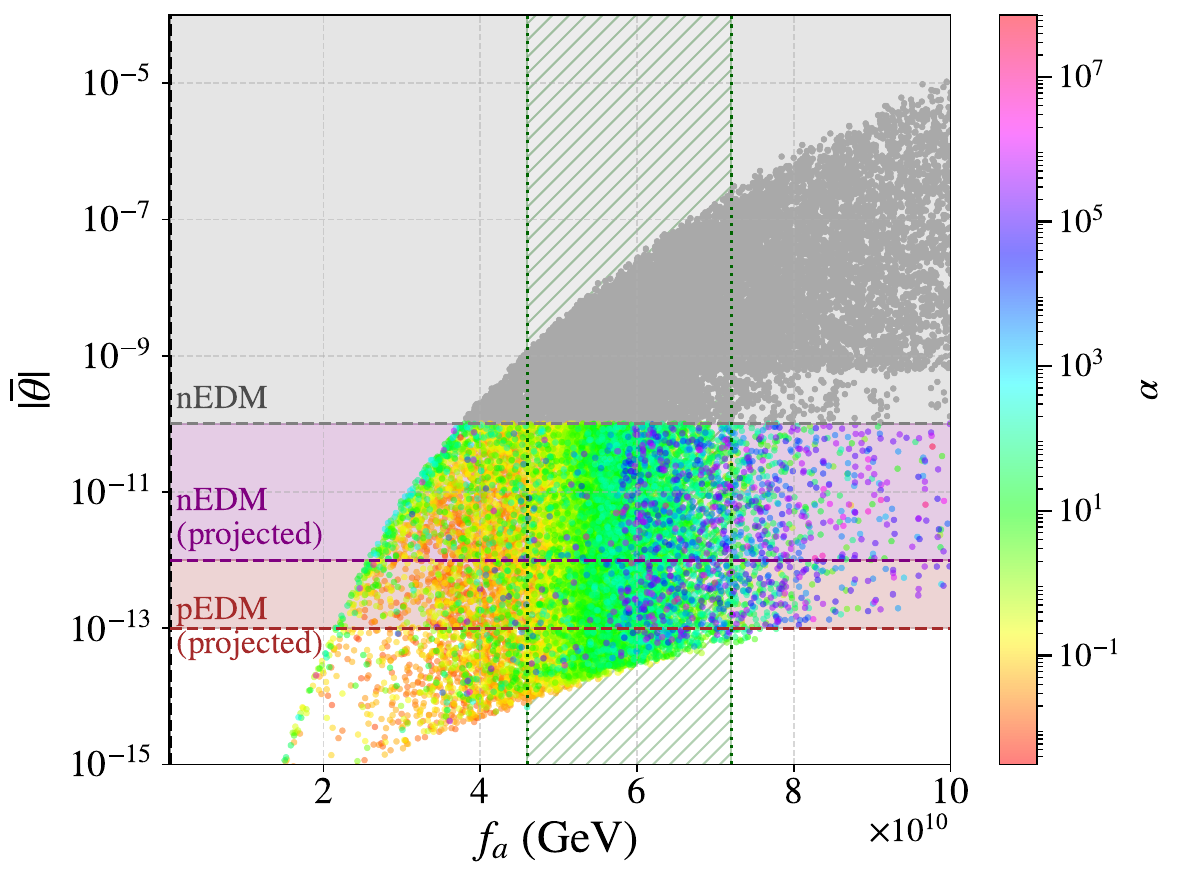}
    \caption{\label{fig: theta_fa_dm} Parameter space in $|\overline{\theta}|$--$f_a$ plane for $n=4$ and $\tilde\mu=7\times 10^{11}\,$GeV. In the band with green lines, axions can explain $100\%$ of dark matter. The grey points are excluded from current nEDM experiments, while the allowed points are colored according to the strength of FOPT via $\alpha$. Also shown are the projected sensitivities from nEDM and pEDM experiments. The entire parameter space depicted lies within the projected sensitivity of searches based on the
    axion–photon coupling.}
\end{figure}

\begin{figure}
    \centering
    \includegraphics[width=0.49\linewidth]{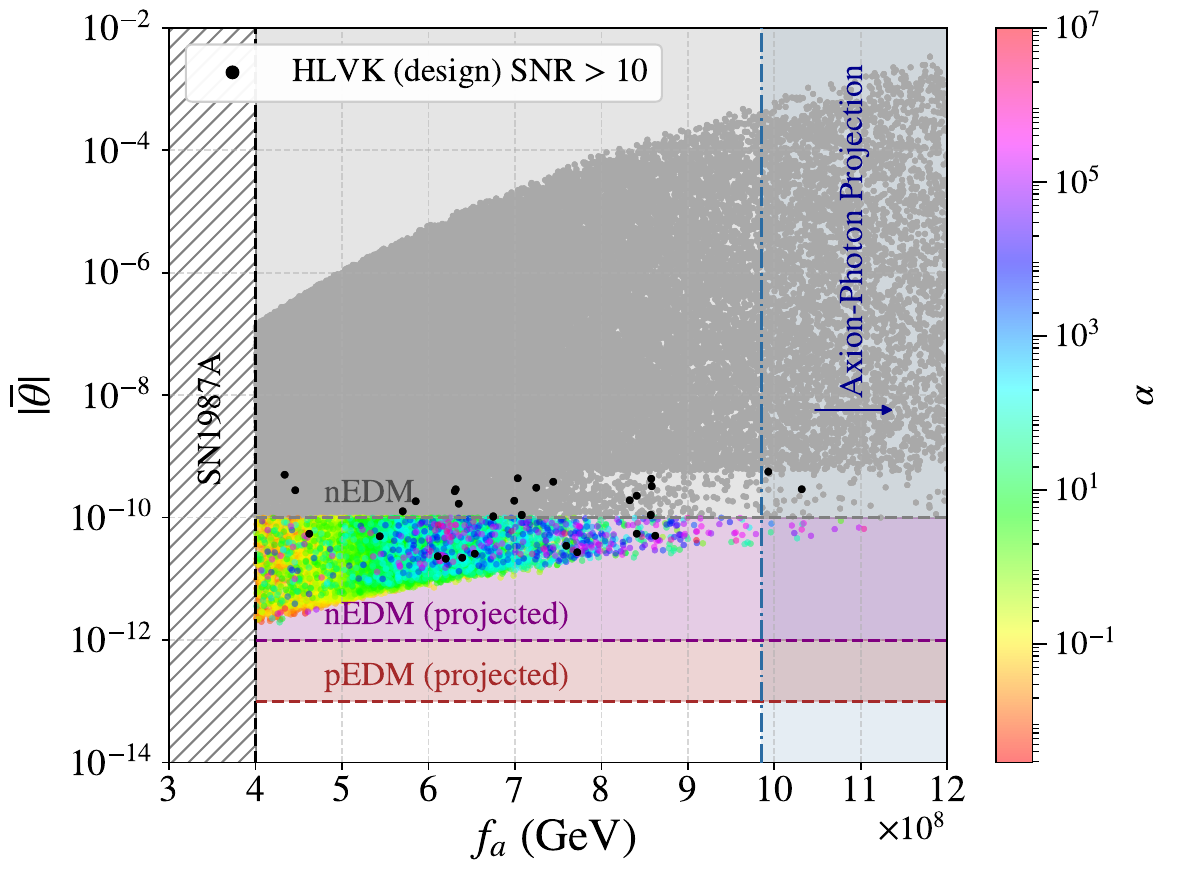}
     \includegraphics[width=0.49\linewidth]{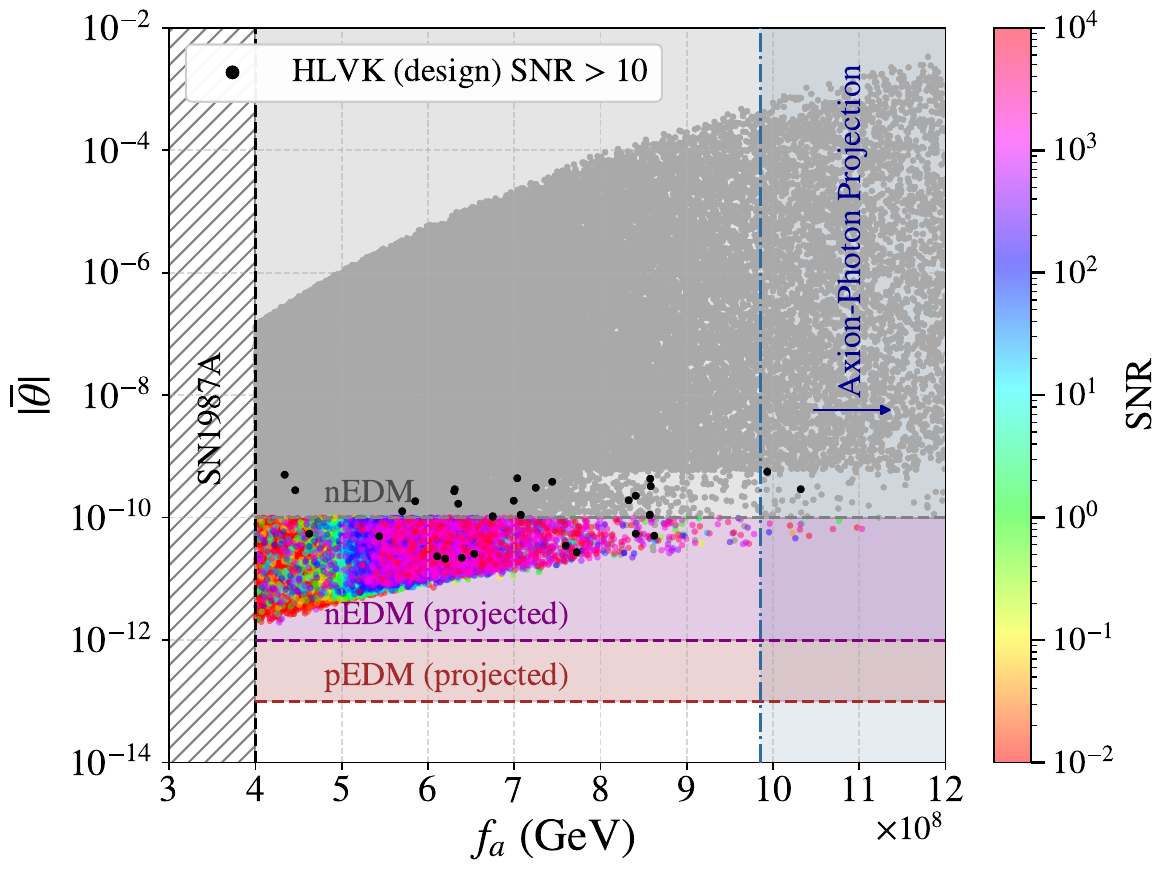}
    \caption{\label{fig: theta_fa} Parameter space in $|\overline{\theta}|$--$f_a$ plane for $n=3$ and $\tilde\mu=5\times 10^{9}\,$GeV. The grey points are excluded from current nEDM bounds, while the allowed points are colored according to the strength of FOPT via $\alpha$ (left panel) and the SNR of the GW signal at the CE detector (right panel). Also shown are the projected sensitivities from future nEDM, pEDM and axion-search experiments. The grey-hatched region towards the left is ruled out due to astrophysical bounds \cite{Raffelt:2006cw, Caputo:2024oqc}.}
\end{figure}



\section{Summary and Conclusion}
\label{sec4}
We summarize our results in Fig. \ref{fig: theta_fa_dm} and Fig. \ref{fig: theta_fa}. The y-axis shows the corrections to QCD vacuum angle, a measure of axion quality while the x-axis shows the axion decay constant. In all these figures, the grey colored points are disfavored from current bound on neutron EDM \cite{Abel:2020pzs}, while the  magneta and red shaded regions denote projected future sensitivities for nEDM \cite{nEDM:2019qgk} and proton EDM (pEDM) \cite{Omarov:2020kws} respectively. While Fig. \ref{fig: theta_fa_dm} considers $n=4$, the other two figures use $n=3$. Smaller $n$ leads to more points violating the nEDM bounds as noted earlier. The color code in Fig. \ref{fig: theta_fa_dm} and the left panel of Fig. \ref{fig: theta_fa} denotes the value of $\alpha$, indicating the strength of the supercooled FOPT. On the other hand, the right panel of Fig. \ref{fig: theta_fa} uses a color code to indicate the SNR at future GW detector CE. Clearly, most of the parameter space allowed by current nEDM bounds can be observed at CE. The blue-shaded region towards the right in Fig. \ref{fig: theta_fa} and the entire parameter space of Fig. \ref{fig: theta_fa_dm} remain within reach of several axion-detection experiments sensitive to axion-photon coupling. In Fig. \ref{fig: theta_fa_dm}, we also show the parameter space consistent with axion being $100\%$ dark matter in the Universe as the green-hatched region. In Fig. \ref{fig: theta_fa}, the black colored points correspond to the parameter space where SNR $>10$ for HLVK operating at design sensitivity. As no statistically significant background has been reported in the current search~\cite{Virgo:2025aai}, these points are likely to be disfavored. Clearly, there exist several complementary detection prospects of the model via observations of stochastic GWs, EDM as well as axion-photon interactions with or without axions being the dark matter in the Universe.

To conclude, we have studied the possibility of probing a quality axion model at future gravitational wave experiments by considering the source of stochastic GWs to be a supercooled first-order phase transition associated with the Peccei-Quinn symmetry breaking. Adopting a specific class of quality axion model based on an additional $U(1)$ gauge symmetry, we study the details of radiative symmetry breaking and a supercooled FOPT by applying relevant phenomenological bounds related to QCD axions and associated particles while being in agreement with the required QCD vacuum angle from neutron EDM searches. After identifying a few benchmark points satisfying these requirements, we perform a numerical scan to determine the allowed parameter space. Using the most up-to-date prescription for calculating GW spectra from supercooled FOPT while including the NLO corrections, we find the prospects of probing the model at current as well as future planned GW experiments. We also calculate the signal-to-noise ratio of the currently allowed parameters for future GW detectors like Cosmic Explorer and find most of them to be within its reach. In fact, a few points in our parameter space also saturates the HLVK limits and are on the verge of being disfavored. Interestingly, most part of this parameter space within reach of future GW experiments can also be observed at future EDM searches while a subset of the parameter space can also be probed at axion haloscope experiments. A part of this observable parameter space can also satisfy the requirement of axion being the dark matter in the Universe. While previous works on GW prospects of quality axion models mostly considered topological defects as the source, we have found equally promising GW signatures via strong supercooled first-order phase transition with complementary detection prospects at future EDM and axion search experiments.

\section*{Acknowledgements}
The work of Debasish Borah is supported by the Science and Engineering Research Board (SERB), Government of India grant CRG/2022/000603. D.R. is supported by the ANRF-NPDF fellowship with grant number PDF/2025/003379.

\appendix
\section{Details of the model with classical conformal invariance}
\label{appen1}
\subsection{Axion Quality Problem}
Using chiral perturbation theory, the zero-temperature QCD axion potential can be derived as 
\begin{equation}
       V \, (a) \, = \,-  f_\pi^2 m_{\pi}^2 \, \sqrt{1 \, - \, \dfrac{4 m_d m_u}{(m_u \, + \, m_d )^2} \, \sin^2 \left(\dfrac{a}{2 f_a} + \dfrac{\bar{\theta}}{2}\right) }.
\end{equation}
At high temperatures, we will get temperature-dependent mass and potential for axion. The $\bar{\theta}$ in the potential can be absorbed by transforming the axion field as $a \rightarrow a + f_a \bar{\theta}$. Using this potential, the axion mass below QCD scale can be found as \cite{Weinberg:1977ma}
\begin{equation*}
     m_a^2 = \dfrac{m_d m_u}{(m_u \, + \, m_d )^2} \dfrac{m_\pi^2 f_\pi^2}{f_a^2} ,   
\end{equation*}
\begin{equation}\label{axion mass}
\implies\,  m_a \, \approx \, 5.7 \, \left(\dfrac{10^{12} \, {\rm GeV}}{f_a}\right)\, {\rm \mu  eV}.
\end{equation}
The potential $V(a)$ is at minimum when
\begin{equation}
    \sin \left(\dfrac{a}{f_a} + \bar{\theta}\right) = 0.
\end{equation}
This minima is shifted when the gravity-induced PQ breaking higher dimensional operators are introduced which only respects the gauge symmetry. The lowest- dimensional PQ breaking but gauge invariant potential involving the scalars $S, \phi$ of our model is of the form:
\begin{equation}
   V_{\rm gravity}= \kappa e^{i \delta}\, \frac{S^{3n}\phi^\dagger}{(3n)! \,M_{\rm Pl}^{3n-3}}+\text{h.c.}
\end{equation}
Using the parametrisation of the scalar fields   $S = \frac{1}{\sqrt{2}}f_S\, e^{i \eta_S/f_S}, \phi =\frac{1}{\sqrt{2}} f_\phi\, e^{i \eta_\phi/f_\phi}$, the corresponding new contribution to the axion potential can be written as, 
\begin{equation}
   V_{\rm gravity}^{(a)}= \kappa\frac{f_S^{3n}f_\phi}{(3n)! \,2^{(3n-1)/2}\,M_{\rm Pl}^{3n-3}} \, \cos \left(\frac{a}{f_a}+\delta\right).
\end{equation}
Here, $ a = \frac{3 n\, f_\phi \eta_S - f_S \eta_\phi}{\sqrt{9 n^2 f_\phi ^2 + f_S^2}}$ and $f_a = \frac{f_S f_\phi}{\sqrt{9 n^2 f_\phi ^2 + f_S^2}}$. After taking into account the gravity-induced correction to the total axion potential which is now a sum of $V_{\rm gravity}^{(a)} + V (a)$ and minimizing it, the total shifted value of the QCD $\bar{\theta}$-vacuum is given as
\begin{equation}
    \bar{\theta} \sim \frac{\kappa \,{\rm sin\delta} \, f_S^{3n}f_\phi}{(3n)! \,2^{(3n-1)/2}\,M_{\rm Pl}^{3n-3} \, m_a^2 f_a^2}.
\end{equation}
However, in addition to the direct contribution to the scalar potential, 4 types of higher dimensional PQ-breaking operators involving the fermion fields. Following Babu et al., the relevant effective Yukawa interactions can be written as
\begin{equation}
    \begin{aligned}
\mathcal{L}_{\rm Yuk}^{\rm gravity}
= \kappa_{ab} \frac{(Q_{aL}Q_{bR})^3\phi}
{M_{\rm Pl}^{6}} + \kappa_{44}
\frac{Q_{4L}Q_{4R}(S^*)^{3n}}
{(3n)!\,M_{\rm Pl}^{3n-1}}+
\kappa_{a4}
\frac{Q_{aL}Q_{4R}(S^*)^n}
{n!,M_{\rm Pl}^{n-1}} + \kappa_{4a}
\frac{Q_{4L}Q_{aR}(S^*)^n}
{n!,M_{\rm Pl}^{n-1}} +\mathrm{h.c.},
\end{aligned}
\end{equation}
where  the $\kappa$ represents the corresponding effective couplings. These interactions can generate additional contributions to the axion potential through loop diagrams.
\\ At the three-loop level, the induced contribution to the scalar potential is estimated as
\begin{equation}
   \begin{aligned}
V_{\rm gravity}^{3\text{-loop}} \sim\kappa_{ab}
\left(\frac{\mathrm{Tr}(Y^{(Q)})}{16\pi^2}\right)^3 \, \frac{S^{3n}\phi^*}
{(3n)! M_*^{3n-3}} \,
\left(\frac{M_*}{M_{\rm Pl}}\right)^6
+\mathrm{h.c.},
\end{aligned} 
\end{equation}
where $Y^{(Q)}$ denotes the relevant Yukawa matrix and $M_*$ is the scale associated with the heavy fermions. The corresponding contribution to $\bar{\theta}$ is obtained by replacing the tree-level coefficient with the loop-induced coefficient in the expression above.\\
There are also one-loop contributions arising from the fermion operators involving VLQ which are given as follows
\begin{equation}
   \begin{aligned}
       V_{\rm gravity}^{1\text{-loop}} \sim \frac{Y_{44}^{(Q)}}{16\pi^2} \left(\kappa_{a4}^* Y_{ab}^{(Q)} \kappa_{4b}^*\right) \,\frac{S^{3n}\phi^*}{(3n)! M_{\rm Pl}^{3n-3}} \left(\frac{M_{\rm Pl}}{M_*}\right)^{n-1} +\mathrm{h.c.},
   \end{aligned}
   \label{loop}
\end{equation}
and \begin{equation}
    \begin{aligned}
        V_{\rm gravity}^{\kappa_{44}}
\sim 
\frac{Y_{44}^{(Q)}}{16\pi^2}\kappa_{44}
\frac{S^{3n}\phi^*}
{(3n)!M_{\rm Pl}^{3n-3}}
\left(\frac{M_*}{M_{\rm Pl}}\right)^2
+\mathrm{h.c.}.
    \end{aligned}
\end{equation}
The total induced strong-CP angle receives contributions from both the direct gravitational operator and the loop-induced terms. We therefore require their combined contribution to satisfy $|\bar{\theta}_{\rm eff}|\lesssim 6\times10^{-10}$.

Among the radiative corrections, the mixed one-loop contribution involving $\kappa_{a4}$ and $\kappa_{4b}$ in Eq. \eqref{loop} can be particularly restrictive because of the enhancement factor $(M_{\rm Pl}/M_*)^{n-1}$. For $n=3$, this enhancement can overcome the loop suppression, depending on the Yukawa couplings and $M_*$. In contrast, the three-loop contribution contains additional loop suppression factors, while the contribution proportional to $\kappa_{44}$ is suppressed by $(M_*/M_{\rm Pl})^2$.

\subsection{Field-dependent masses in the conformal model}
\label{app: masses}
\subsubsection{Masses in the $s-\varphi$ plane}
For scalars, the field dependent masses are computed using
\[[\mathcal{M}_{\rm{scalar}}^2]_{ij}=\frac{\partial^2V_0}{\partial\phi_i\partial\phi_j}\]
where \(\phi_i\) represents the scalar fields in the model.
For $\phi_i\in\left\{s,\varphi,\xi,\eta\right\}$, the $4\times 4$ scalar mass matrix is given by

\begin{equation} 
\mathcal{M}_{\rm{scalar}}^2 = \begin{pmatrix} 
m^2_s & m^2_{s\varphi} & 0 & 0 \\[1ex] 
m^2_{s \varphi} & m^2_\varphi & 0 & 0 \\[1ex] 
0 & 0 & m^2_\xi & 0 \\[1ex] 
0 & 0 & 0 & m^2_\eta
\end{pmatrix}, \label{eq: mscalar} 
\end{equation}
where
\begin{align*}
m^2_s &= 3\lambda_S s^2 + \frac{\lambda_{\phi S}}{2}\varphi^2, \,\, m^2_{s\varphi} = \lambda_{\phi S}s\varphi, \,\, m^2_\varphi = 3\lambda_\phi \varphi^2 + \frac{\lambda_{\phi S}}{2}s^2, \\
m^2_\xi &= \lambda_S s^2 + \frac{\lambda_{\phi S}}{2}\varphi^2, \,\,
m^2_\eta = \lambda_\phi \varphi^2 + \frac{\lambda_{\phi S}}{2}s^2.
\end{align*}
The fermion masses are obtained from the Yukawa Lagrangian given in Eq. \eqref{eq: L_yuk}. The masses of quarks $Q_{1,2,3}$ are generated by higher dimensional operators as
\beq
\label{eq: mQ123}
M_{Q, ab} = Y^{(Q)}_{ab}\frac{s^n}{2^{n/2}n!M_*^{n-1}},\quad a,b\in\{1,2,3\}
\eeq
Assuming a diagonal Yukawa matrix $Y^{(Q)}$, the squared mass matrix, $\mathcal{M}^2_{Q} \equiv M_Q^{\dagger} M_Q$, is given by
\beq
\label{eq: mQ123}
\mathcal{M}^2_{Q} = \frac{M_*^2}{2^{n}(n!)^2}\left(\frac{s}{M_*}\right)^{2n}\,{\rm{diag}}\left(|Y^{(Q)}_{11}|^2,\,|Y^{(Q)}_{22}|^2,\,|Y^{(Q)}_{33}|^2\right)
\eeq
The remaining quark, $Q_4$, has field-dependent mass
\beq
\label{eq: mQ4}
m^2_{Q_4} = \frac{1}{2}|Y^{(Q)}_{44}|^2\varphi^2\,.
\eeq
The right handed neutrino masses are given by
\begin{align}
\label{eq: mN12}
m^2_{N_{1,2}} &= \frac{1}{2}|Y^{(N)}_{12}|^2\varphi^2\nn\\
m^2_{N_3} &= \frac{M_*^2}{16}\left(\frac{\varphi}{M_*}\right)^4 |Y^{(N)}_{33}|^2
\end{align}
Lastly, the field dependent mass for the gauge boson associated with $U(1)_a$ is given by
\begin{equation}
    \label{eq: mX}
m_X^2
=
g_X^2\left(
q_S^2\,s^2
+
q_\phi^2\,\varphi^2
\right), \quad q_S=\frac{2}{n},\,q_\phi = 6,\,n=4.
\end{equation}
Next, we list the thermal Debye masses. The scalar thermal masses are: 
\begin{align}
\label{eq: Pi_scalar}
    \Pi_{s}(T) &=\frac{T^2}{12}\left(\frac{12}{n^2}\,g_X^2+4\lambda_S+\lambda_{\phi S}+\frac{6n(2n-1)}{2^n(n!)^2}\left(\frac{s}{M_*}\right)^{2n-2} \,\sum_{i=1}^3 |Y^{(Q)}_{ii}|^2\right)\,, \\
    \Pi_{\varphi}(T) &=\frac{T^2}{24}\left(216\,g_X^2+2(Y^\mathrm{(N)}_{12})^2+6(Y^\mathrm{(Q)}_{44})^2+8\lambda_\phi+2\lambda_{\phi S}+\frac{3}{4}\left(\frac{\varphi}{M_*}\right)^2 |Y^{(N)}_{33}|^2 \right)\,,\\
    \Pi_{\xi}(T) &=\frac{T^2}{12}\left(\frac{12}{n^2}\,g_X^2+4\lambda_S+\lambda_{\phi S}+\frac{6n}{2^n(n!)^2}\left(\frac{s}{M_*}\right)^{2n-2} \,\sum_{i=1}^3 |Y^{(Q)}_{ii}|^2\right)\,, \\
    \Pi_{\eta}(T) &=\frac{T^2}{24}\left(216\,g_X^2+2(Y^\mathrm{(N)}_{12})^2+6(Y^\mathrm{(Q)}_{44})^2+8\lambda_\phi+2\lambda_{\phi S}+\frac{1}{4}\left(\frac{\varphi}{M_*}\right)^2 |Y^{(N)}_{33}|^2 \right)\,.
\end{align}
The thermal mass for the longitudinal component of the gauge boson is
\begin{align}
\label{eq: Pi_gauge}
    \Pi_X(T)&=g_X^2T^2\left(\frac{1}{3}\sum_{\rm{complex~ scalars}}q_s^2+\frac{1}{6}\sum_{\rm{Weyl~ fermions}}q_f^2\right)\nn\\
    &=g_X^2 T^2\left(\frac{100}{3}+\frac{4}{3n^2}\right).
\end{align}
The thermally corrected masses used in Eq. \eqref{eq: Vdaisy} are eigenvalues of the respective mass matrices obtained by replacing
$$\mathcal{M}^2\rightarrow \mathcal{M}^2 + \Pi.$$

\subsubsection{Masses along the flat direction}
\label{app: masses_flat}
The field-dependent masses in polar coordinates can be obtained using Eq. \eqref{eq: polar} in the expressions for $m_i^2(\varphi, s)$ given above. Flat direction masses are then obtained by setting $\Theta= \widetilde\Theta$. The scalar matrix becomes \footnote{Here, all couplings are taken at $\mu=\tilde\mu$. The tilde is dropped for brevity.}
\beq
\mathcal{M}_{\rm{scalar}}^2=\begin{pmatrix}
\displaystyle
\frac{2r^2\lambda_\phi\left(\lambda_s+\sqrt{\lambda_s\lambda_\phi}\right)}
{D}
&
\displaystyle
-\frac{2r^2\lambda_s\lambda_\phi}
{\sqrt{\lambda_s+\sqrt{\lambda_s\lambda_\phi}}\,
 \sqrt{\lambda_\phi+\sqrt{\lambda_s\lambda_\phi}}}
&
0 & 0
\\[12pt]
\displaystyle
-\frac{2r^2\lambda_s\lambda_\phi}
{\sqrt{\lambda_s+\sqrt{\lambda_s\lambda_\phi}}\,
 \sqrt{\lambda_\phi+\sqrt{\lambda_s\lambda_\phi}}}
&
\displaystyle
\frac{2r^2\lambda_s\left(\lambda_\phi+\sqrt{\lambda_s\lambda_\phi}\right)}
{D}
&
0 & 0
\\[12pt]
0 & 0 & 0 & 0
\\
0 & 0 & 0 & 0
\end{pmatrix}\,,
\eeq
where $D \equiv \lambda_S + \lambda_\phi + 2\sqrt{\lambda_S\lambda_\phi}$. Diagonalizing this matrix yields one massive field $s_\perp$ and three massless fields corresponding to two goldstones and one scalon. 
\beq
[\mathcal{M}_{\rm{scalar}}^2]^{\rm{diag}} = {\rm{diag}}\left(m_{s_\perp}^2,0,0,0\right)\,,
\eeq
where the non-zero entry is
\beq
m_{s_\perp}^2 = \kappa_{s_\perp} r^2\,, \quad \kappa_{s_\perp} =  2\sqrt{\lambda_S\lambda_\phi}\,.
\eeq 
Similarly, the renormalizable masses of the gauge boson and fermions $N_{1,2}$ and $Q_4$ have the form
\beq
m^2_{N_{1,2}} = \kappa_{N}r^2\,,~m^2_{Q_{4}} = \kappa_{Q}r^2\,,~m^2_{X} = \kappa_{X}r^2\,.
\eeq
The $\kappa$s are given by
\begin{align}
\kappa_N &=\frac{|Y^{(N)}_{12}|^2\left(\lambda_S + \sqrt{\lambda_S\lambda_\phi}\right)}{2D}\,,\nn\\
\kappa_Q &=\frac{|Y^{(Q)}_{44}|^2\left(\lambda_S + \sqrt{\lambda_S\lambda_\phi}\right)}{2D}\,,\nn\\
\kappa_X &= \frac{g_X^2\left(144 \lambda_S + \lambda_\phi + 145\sqrt{\lambda_S\lambda_\phi}\right)}{4D}\,.
\end{align} 
For the non-renormalizable masses, $m^2_{N_3} = \kappa_{N_3}r^4$, $m^2_{Q_i} = \kappa_{Q_i}r^8$, we have
\begin{align}
    \kappa_{N_3} &= \frac{|Y^{(N)}_{33}|^2\lambda_S}{16 M_*^2 D}\,,\nn\\
    \kappa_{Q_i} &= \frac{|Y^{(Q)}_{ii}|^2(\lambda_\phi + \sqrt{\lambda_S\lambda_\phi})^4}{9216 M_*^6 D^4}\,,~i\in\{1,2,3\}.
\end{align}

\subsection{Beta functions}
\label{app: beta}

The one-loop beta functions for the renormalizable operators, defined via $\frac{1}{16\pi^2}\beta_{\lambda}^{(1)}\equiv \mu\frac{d\lambda}{d\mu}$, have been computed using the \texttt{SARAH} package \cite{Staub:2008uz}. Here we present the expressions for a general $n$ which appears in the $U(1)_a$ charge of $S$ as $q_S=2/n$. The one-loop beta functions for the relevant model parameters are given as follows.
%
%
\begin{align}
\beta_{g_1}^{(1)}
={}&
\frac{251}{30}g_1^3, \quad \beta_{g_2}^{(1)}
={}
-\frac{19}{6}g_2^3, \quad \beta_{g_3}^{(1)}
={}
-\frac{13}{3}g_3^3, \quad \beta_{g_X}^{(1)}
={}
\left(
\frac{292}{3}
+\frac{4}{3n^2}
\right)g_X^3,
\label{eq:beta_g1}
\end{align}
\begin{align}
\beta_{y_t}^{(1)}
&=
y_t
\left[
\frac{9}{2}y_t^2
-\frac{17}{20}g_1^2
-\frac{9}{4}g_2^2
-8g_3^2
\right],
\\[4pt]
\beta_{Y^{(N)}_{12}}^{(1)}
&=
Y^{(N)}_{12}
\left[
2\left(Y^{(N)}_{12}\right)^2
+3\left(Y^{(Q)}_{44}\right)^2
-60g_X^2
\right],
\\[4pt]
\beta_{Y^{(Q)}_{44}}^{(1)}
&=
Y^{(Q)}_{44}
\left[
\left(Y^{(N)}_{12}\right)^2
+4\left(Y^{(Q)}_{44}\right)^2
-\frac{8}{5}g_1^2
-8g_3^2
-54g_X^2
\right].
\end{align}
\begin{align}
\beta_{\lambda_H}^{(1)}
={}&
\frac{27}{200}g_1^4
+\frac{9}{20}g_1^2g_2^2
+\frac{9}{8}g_2^4
-\frac{9}{5}g_1^2\lambda_H
-9g_2^2\lambda_H
\nonumber\\
&+
24\lambda_H^2
+\lambda_{HS}^2
+\lambda_{H\phi}^2
+12\lambda_H y_t^2
-6y_t^4,
\label{eq:beta_lamH}
\\[2mm]
\beta_{\lambda_{HS}}^{(1)}
={}&
-\frac{9}{10}g_1^2\lambda_{HS}
-\frac{9}{2}g_2^2\lambda_{HS}
-\frac{24}{n^2}g_X^2\lambda_{HS}
\nonumber\\
&+
12\lambda_H\lambda_{HS}
+4\lambda_{HS}^2
+8\lambda_{HS}\lambda_S
+2\lambda_{H\phi}\lambda_{\phi S}
\nonumber\\
&+6\lambda_{HS}y_t^2,
\label{eq:beta_lamHS}
\\[2mm]
\beta_{\lambda_{H\phi}}^{(1)}
={}&
-\frac{9}{10}g_1^2\lambda_{H\phi}
-\frac{9}{2}g_2^2\lambda_{H\phi}
-216g_X^2\lambda_{H\phi}
\nonumber\\
&+
12\lambda_H\lambda_{H\phi}
+4\lambda_{H\phi}^2
+2\lambda_{HS}\lambda_{\phi S}
+8\lambda_{H\phi}\lambda_\phi
\nonumber\\
&+
2\lambda_{H\phi}\left(Y^{(N)}_{12}\right)^2
+6\lambda_{H\phi}\left(Y^{(Q)}_{44}\right)^2
+6\lambda_{H\phi}y_t^2,
\label{eq:beta_lamHT}
\\[2mm]
\beta_{\lambda_S}^{(1)}
={}&
\frac{96}{n^4}g_X^4
+2\lambda_{HS}^2
-\frac{48}{n^2}g_X^2\lambda_S
+20\lambda_S^2
+\lambda_{\phi S}^2,
\label{eq:beta_lamS}
\\[2mm]
\beta_{\lambda_{\phi S}}^{(1)}
={}&
\frac{1728}{n^2}g_X^4
+4\lambda_{HS}\lambda_{H\phi}
-\left(
216+\frac{24}{n^2}
\right)g_X^2\lambda_{\phi S}
\nonumber\\
&+
8\lambda_S\lambda_{\phi S}
+4\lambda_{\phi S}^2
+8\lambda_{\phi S}\lambda_\phi
+2\lambda_{\phi S}\left(Y^{(N)}_{12}\right)^2
+6\lambda_{\phi S}\left(Y^{(Q)}_{44}\right)^2,
\label{eq:beta_lamST}
\\[2mm]
\beta_{\lambda_\phi}^{(1)}
={}&
7776g_X^4
+2\lambda_{H\phi}^2
+\lambda_{\phi S}^2
-432g_X^2\lambda_\phi
+20\lambda_\phi^2
\nonumber\\
&+
4\lambda_\phi \left(Y^{(N)}_{12}\right)^2
-2\left(Y^{(N)}_{12}\right)^4
+12\lambda_\phi \left(Y^{(Q)}_{44}\right)^2
-6\left(Y^{(Q)}_{44}\right)^4.
\label{eq:beta_lamT}
\end{align}
Although we have listed the beta functions for the important SM couplings as well as cross-quartic couplings of new scalars with the SM Higgs doublet for completeness, they are not much relevant for the discussion of radiative breaking of PQ symmetry at high scale.

\section{Gravitational Waves from Supercooled FOPT}
\label{appen1b}
As mentioned in the main text, the total contribution to the stochastic GW background redshifted to today is given by
\begin{equation}
\label{eqGW1}
h^2 \Omega_{\mathrm{GW},0}
= \mathcal{R} \left(
h^2 \Omega_{\mathrm{col}}
+ h^2 \Omega_{\mathrm{sw}}
+ h^2 \Omega_{\mathrm{turb}}
\right),
\end{equation}
where we have the redshift factor for the GW amplitude as
\begin{equation}
\mathcal{R} = \Omega_{\gamma}
\left( \frac{g_{s,0}}{g_{s,\mathrm{reh}}} \right)^{4/3}
\frac{g_{*,\mathrm{reh}}}{g_{*,0}}.
\label{eqGW2}
\end{equation}
In the above expressions, \(\Omega_\gamma=2.473\times10^{-5}/h^2\) denotes the photon energy density, \(g_{s,0}\) and  \(g_{\ast,0}\) are the effective degrees of freedom contributing to the entropy and energy density of the Universe at the present epoch. \(g_{s,{\rm reh}}\) and  \(g_{\ast,{\rm reh}}\) are the same quantities evaluated at the time of reheating, when FOPT transfers all its energy to the radiation bath of the Universe. We also consider the Hubble parameter at the \(H_\ast\) at the time of formation of GW, redshifted to today as
\begin{equation}
H_{*,0} = \frac{a_*}{a_0} H_*
\simeq 1.65 \times 10^{-5}\,\mathrm{Hz}
\left( \frac{g_*}{100} \right)^{1/6}
\left( \frac{T_*}{100\,\mathrm{GeV}} \right),
\label{eqGW3}
\end{equation}
where $a$ is the scale factor. The bubble wall contribution to GW is given by
\begin{equation}
h^2 \Omega_{\mathrm{col}}(f)
= h^2 A_{\mathrm{col}} K_{\mathrm{col}}^2
\left[\frac{\beta}{H_{p}} \right]^2
S_{\mathrm{col}}(f),
\qquad
A_{\mathrm{col}} = 0.05,
\label{eqGW4}
\end{equation}
where the subscript $p$ denotes the quantity being evaluated at the percolation temperature $T_p$. The kinetic energy fraction and the spectral shape of the bubble wall collision are given by
\begin{equation}
K_{\mathrm{col}} = \kappa_{\mathrm{col}} \frac{\alpha}{1+\alpha},
\qquad
S_{\mathrm{col}}(f) =
\left( \frac{f}{f_{\mathrm{col}}} \right)^{2.4}
\left[
\frac{1}{2} + \frac{1}{2}
\left( \frac{f}{f_{\mathrm{col}}} \right)^{1.2}
\right]^{-4},
\qquad
f_{\mathrm{col}} \simeq 0.11
{H_{*,0}}\frac{\beta}{H_\ast}.
\label{eqGW5}
\end{equation}
The sound wave contribution is given by
\begin{equation}
h^2 \Omega_{\mathrm{sw}}(f)
= h^2 A_{\mathrm{sw}} K_{\mathrm{sw}}^2
\left[ R_p H(T_p) \right]
\Upsilon_{\mathrm{sw}} S_{\mathrm{sw}}(f),
\qquad
A_{\mathrm{sw}} \simeq 0.11,
\label{eqGW6}
\end{equation}
Here, we can use the following relation to have the formulae in terms of \(\beta/H_\ast\),
\begin{equation}
\left( \frac{\beta}{H} \right)_{p}
=
\left( {8\pi} \right)^{1/3}
\frac{\max(v_w,c_s)}{R_pH(T_p)}
\label{eqGW7}
\end{equation}
The kinetic energy fraction for the sound waves, and the spectral shape are given by
\begin{equation}
K_{\mathrm{sw}} = 0.6 \kappa_{\mathrm{sw}} \frac{\alpha}{1+\alpha},
\qquad
S_{\mathrm{sw}}(f) = N
\left( \frac{f}{f_{\mathrm{sw},1}} \right)^3
\left[
1 + \left( \frac{f}{f_{\mathrm{sw},1}} \right)^2
\right]^{-1}
\left[
1 + \left( \frac{f}{f_{\mathrm{sw},2}} \right)^4
\right]^{-1},
\label{eqGW8}
\end{equation}
where
\begin{equation}
f_{\mathrm{sw},1} \simeq 0.2
\frac{H_{*,0}}{R_p H(T_p)},
\qquad
f_{\mathrm{sw},2} \simeq
\frac{0.5}{\Delta_w}
\frac{H_{*,0}}{R_p H(T_p)},
\qquad
\Delta_w =
\frac{|v_w-c_s|}{\max(v_w,c_s)}.
\label{eqGW9}
\end{equation}
The factor \(N\) comes from the normalisation condition
\begin{equation}
\int_{-\infty}^{\infty} S_{\mathrm{sw}}(f)\, d\ln f = 1
\Longrightarrow
N = \frac{4}{\pi}
\frac{
f_{\mathrm{sw},1}
\left(f_{\mathrm{sw},1}^{4} + f_{\mathrm{sw},2}^{4}\right)
}{
f_{\mathrm{sw},2}^{3}
\left(
\sqrt{2}f_{\mathrm{sw},1}^{2}
- 2f_{\mathrm{sw},1}f_{\mathrm{sw},2}
+ \sqrt{2}f_{\mathrm{sw},2}^{2}
\right)
},
\label{eqGW10}
\end{equation}
whereas \(\Upsilon_{\rm sw}\) is the suppression factor in a radiation dominated Universe, given by~\cite{Guo:2020grp}
\begin{equation}
\Upsilon_{\mathrm{sw}} =
1 - \frac{1}{\sqrt{2\tau_{\mathrm{sh}}H(T_p)+1}},
\qquad
\tau_{\mathrm{sh}} \equiv
\frac{2R_p}{\sqrt{3K_{\mathrm{sw}}}}.
\label{eqGW11}
\end{equation}
Finally, the contribution from turbulence is given by
\begin{equation}
h^2 \Omega_{\mathrm{turb}}(f)
= h^2 A_{\mathrm{turb}} K_{\mathrm{turb}}^2
\left[ R_p H(T_p) \right]^3
S_{\mathrm{turb}}(f),
\qquad
A_{\mathrm{turb}} \simeq 0.255,
\label{eqGW12}
\end{equation}
with the kinetic energy fraction and the spectral shape as
\begin{equation}
K_{\mathrm{turb}} = 0.6 \kappa_{\mathrm{turb}} \frac{\alpha}{1+\alpha},
\qquad
\kappa_{\mathrm{turb}} = \epsilon \kappa_{\mathrm{sw}},
\qquad
\epsilon = 0.1,
\label{eqGW13}
\end{equation}
\begin{equation}
S_{\mathrm{turb}}(f) =
\left( \frac{f}{f_{\mathrm{turb},1}} \right)^3
\left[
1 + \left( \frac{f}{f_{\mathrm{turb},2}} \right)^{2.15}
\right]^{-7.9}
\begin{cases}
\ln^2 \left( 1 + \dfrac{H_{*,0}}{2\pi f_{\mathrm{turb},3}} \right),
& \text{for } f \leq f_{\mathrm{turb},3}, \\[8pt]
\ln^2 \left( 1 + \dfrac{H_{*,0}}{2\pi f} \right),
& \text{for } f > f_{\mathrm{turb},3},
\end{cases}
\label{eqGW14}
\end{equation}
where,
\begin{equation}
f_{\mathrm{turb},1} = H_{*,0},
\qquad
f_{\mathrm{turb},2} = 2.2
\frac{H_{*,0}}{R_p H(T_p)},
\qquad
f_{\mathrm{turb},3} =
\frac{\sqrt{3K_{\mathrm{turb}}}}{4}
\frac{H_{*,0}}{R_p H(T_p)}.
\label{eqGW15}
\end{equation}

The distribution of the energy released during a first-order phase transition among the different gravitational-wave sources depends on the dynamics of the expanding bubble walls. In particular, the bubble-wall velocity \(v_w\) plays an important role in determining the efficiency with which the released vacuum energy is transferred into scalar-field gradients and bulk plasma motion. The local thermal equilibrium approximation gives  a upper bound on the wall velocity \(v_w \leq v_w^{\rm LTE}\).~\cite{Ai:2023see} A precise determination of \(v_w\) requires solving the coupled Boltzmann equations for the particle species interacting with the bubble wall, which is computationally demanding. Given the strongly supercooled nature of the phase transitions considered in this work, we take the velocity of bubble wall, \(v_w\rightarrow 1\) throughout our analysis.

The acceleration of the bubble wall is governed by the competition between the driving pressure arising from the vacuum energy difference and the friction exerted by the surrounding plasma. The leading-order (LO) friction approaches a constant  in the ultra-relativistic limit independent of the wall Lorentz factor \(\gamma_w\), commonly referred to as the Bödeker--Moore friction~\cite{Bodeker:2009qy}. This contribution arises from the changes in the effective masses of particles across the bubble wall and can be expressed as
\begin{equation}
    P_{\mathrm{LO}}
=
\frac{T_p^2}{24}
\sum_{i\in B} n_i\Delta m_i^2
+
\frac{T_p^2}{48}
\sum_{i\in F} n_i\Delta m_i^2,
\label{eqGW17}
\end{equation}

where \(n_i\) denotes the internal degrees of freedom of the particle species \(i\), and \(\Delta m_i^2\) represents the change in its squared mass between the true and false vacua. The sums extend over the bosonic and fermionic degrees of freedom, respectively. The corresponding critical transition strength is defined as

\begin{equation}
    \label{eqGW18}
    \alpha_\infty
=
\frac{P_{\mathrm{LO}}}{\rho^{\rm false}_{\mathrm{rad}}(T_p)},
\end{equation}
where \(\rho_{\mathrm{rad}}(T_p)\) denotes the radiation energy density at the percolation temperature.

For our analysis, we adopt the next-to-leading-order (NLO)
friction prescription, following Refs.~\cite{Espinosa:2010hh,Ellis:2019oqb}. In this
prescription, gauge-boson splitting contributes a friction pressure
proportional to $\gamma_w$. Its coefficient and the resulting terminal
Lorentz factor are
\begin{equation}
    \alpha_{\mathrm{eq}}
    =
    \frac{T_p^3}{\rho_{\mathrm{rad}}^{\mathrm{false}}(T_p)}
    \sum_{i\in\mathrm{gauge}} n_i g_i^2 \Delta m_i,
    \qquad
    \gamma_{\mathrm{eq}}
    =
    \frac{\alpha-\alpha_\infty}{\alpha_{\mathrm{eq}}},
    \label{eq:GW-NLO-friction}
\end{equation}
where $g_i$ is the relevant gauge coupling and
$\Delta m_i=m_{i,\mathrm{true}}-m_{i,\mathrm{false}}$.
We compare $\gamma_{\mathrm{eq}}$ with the maximally possible wall boost factor which can be attained at the time of collision, $\gamma_*=2R_{\mathrm{sep}}/(3R_0)$.
Here $R_{\mathrm{sep}}\simeq R_p$ is the mean bubble separation at
percolation and $R_0$ is the critical radius at bubble formation.
The fraction of released vacuum energy retained in the walls at
collision is
\begin{equation}
    \kappa_{\mathrm{col}}
    =
    \begin{cases}
        1-\dfrac{\alpha_\infty}{\alpha},
        & \gamma_*<\gamma_{\mathrm{eq}},
        \\[6pt]
        \dfrac{\gamma_{\mathrm{eq}}}{\gamma_*}
        \left[
            1-\dfrac{\alpha_\infty}{\alpha}
            \left(\dfrac{\gamma_{\mathrm{eq}}}{\gamma_*}\right)^2
        \right],
        & \gamma_{\mathrm{eq}}\leq\gamma_*
          \quad\text{(relativistic terminal walls)},
        \\[6pt]
        0, & \text{nonrelativistic walls}.
    \end{cases}
    \label{eqGW19}
\end{equation}
In the analysis of our benchmark points, we see that our strongly supercooled phase transitions involving gauge bosons give \(\gamma_{\rm eq}\ll \gamma_\ast\) and \(\alpha \gg \alpha_{\rm eq}\). This leads to \(\kappa_{\rm col}\rightarrow0\), even if the bubble walls were relativistic at the time of the collision. The fraction $1-\kappa_{\mathrm{col}}$ of the released vacuum energy
is transferred to the plasma as bulk motion and heat. The efficiency
for generating bulk fluid motion can be expressed in terms of the
enthalpy density $w(\xi)$ and fluid velocity profile $v(\xi)$,
\begin{equation}
    \kappa_{\mathrm{sw}}
    =
    \frac{3}{v_w^3}
    \int \mathrm{d}\xi\,\xi^2
    \frac{v^2\,w}{1-v^2},
    \label{eqGW20}
\end{equation}
where $\xi=r/t$.
To evaluate this efficiency, we use the fitting functions given in
the appendix of Ref.~\cite{Espinosa:2010hh} in the $v_w\to1$ limit. Lastly, for the turbulence efficiency, we take the efficiency factor for the turbulence as some fraction \(\epsilon_{\rm turb}\) of the sound wave efficiency factor~\cite{Caprini:2024hue}
\begin{equation}
    \label{eqGW21}
    \kappa_{\rm turb} = 0.1\,\kappa_{\rm sw},
\end{equation}
with \(\kappa_{\rm turb} = 0.1\).


In order to project the entire parameter space against the sensitivities of GW detectors, we calculate the SNR as~\cite{Allen:1997ad,Maggiore:1999vm}
\begin{equation}
\mathrm{SNR} = \rho =
\left[
\delta \times t_{\mathrm{obs}}
\int_{f_{\min}}^{f_{\max}}
\frac{df}{\mathrm{Hz}}
\left(
\frac{\Omega_{\mathrm{GW}}h^2(f)}
{\Omega_{\mathrm{noise}}h^2(f)}
\right)^2
\right]^{1/2}.
\label{eqGW16}
\end{equation}
Here, \(\delta\) accounts for the correlation scheme employed in searching for a stochastic gravitational-wave background, with \(\delta=1\) for auto-correlation and \(\delta=2\) for cross-correlation measurements. Accordingly, we adopt \(\delta=1\) for CE and \(\delta=2\) for ET, HLVK~\cite{Schmitz:2020syl}. The parameter \(t_{\mathrm{obs}}\) represents the total observation time of the experiment, which we take to be five years in our analysis. The quantity \(h^2\Omega_{\mathrm{noise}}(f)\) in the denominator denotes the effective noise spectrum of the detector, while \(h^2\Omega_{\mathrm{GW}}(f)\) in the numerator represents the predicted gravitational-wave signal from the model under consideration. The detectability of the resulting stochastic gravitational-wave background is then assessed through the corresponding SNR. To assess the detection prospects, we adopt a threshold of $(\mathrm{SNR}>10)$ for the detectors CE, ET and the HLVK network.





\section{Non-conformal Model}
\label{appen2}
The tree-level scalar potential is
\begin{equation}
    V_\mathrm{tree} = -\mu_S^2 \left|S\right|^2-\mu_\phi^2 \left|\phi\right|^2+\lambda_S \left(S^{\dagger}S\right)^2+\lambda_\phi \left(\phi^{\dagger}\phi\right)^2+\lambda_{\phi S}\left(S^{\dagger}S\right)\left(\phi^{\dagger}\phi\right)
    \label{eq2}
\end{equation}
Only the real parts of these fields get vev, hence the tree level potential in terms of the real parts is:
\begin{equation*}
    V_\mathrm{tree,real}=-\mu_S^2\frac{s^2}{2}-\mu_\phi^2\frac{\varphi^2}{2}+\lambda_S\frac{s^4}{4}+\lambda_\phi\frac{\varphi^4}{4}+\lambda_{\phi S}\frac{s^2\varphi^2}{4}.
\end{equation*}
We keep the vevs of \(\phi\) and \(S\) to be at \(f_\phi\) and \(f_S\) respectively, which we get by minimizing this tree potential:
\begin{equation}
    f_\phi=\frac{\sqrt{2}\sqrt{2 \lambda_S \mu_\phi^2-\lambda_{\phi S}\mu_S^2}}{\sqrt{4\lambda_S \lambda_\phi-\lambda_{\phi S}^2}},\quad\quad f_S=\frac{\sqrt{2}\sqrt{2 \lambda_\phi \mu_S^2-\lambda_{\phi S}\mu_\phi^2}}{\sqrt{4\lambda_S \lambda_\phi-\lambda_{\phi S}^2}}.
    \label{eq3}
\end{equation}
This forces the condition 
\[
4 \lambda_S \lambda_\phi > \lambda^2_{\phi S}
\]
to hold, so that we can have valid vevs for both scalars. Inverting these relations will give us the \(\mu_S^2\) and  \(\mu_\phi^2\) in terms of vevs,
\begin{equation*}
    \mu_\phi^2=f_\phi^2\lambda_\phi+\frac{1}{2}f_S^2\lambda_{\phi S},\quad\quad \mu_S^2=f_S^2\lambda_S+\frac{1}{2}f_\phi^2\lambda_{\phi S}
\end{equation*}

The field dependent masses can be found by using
\[\frac{\partial^2V}{\partial\phi_i\partial\phi_j}\]
where \(\phi_i\)'s are the species in the model. In our model, the sequence of the scalar fields are \(\left(\varphi,s,\eta,\xi\right)\), which gives us the corresponding mass matrix:
\begin{equation}
\mathcal{M}_{ij}^2 =
\begin{pmatrix}
3\lambda_\phi \varphi^2 + \dfrac{\lambda_{\phi S}}{2}s^2 - \mu_\phi^2
&
\lambda_{\phi S}s\varphi
&
0
&
0
\\[2ex]
\lambda_{\phi S}s\varphi
&
3\lambda_S s^2 + \dfrac{\lambda_{\phi S}}{2}\varphi^2 - \mu_S^2
&
0
&
0
\\[2ex]
0
&
0
&
\lambda_\phi \varphi^2 + \dfrac{\lambda_{\phi S}}{2}s^2 - \mu_\phi^2
&
0
\\[2ex]
0
&
0
&
0
&
\lambda_S s^2 + \dfrac{\lambda_{\phi S}}{2}\varphi^2 - \mu_S^2
\end{pmatrix}.
\label{eq9}
\end{equation}
Diagonalising this mass matrix will give us the physical field dependent masses of the scalar fields. The Yukawa sector gives us the field dependent masses of the fermions which couple to \(\phi\) and \(S\). The right handed fermion sector have their field dependent masses as:
\begin{equation}
M_N^2 =
\begin{pmatrix}
\dfrac{(Y^{(N)}_{12})^2\,\varphi^2}{2} & 0 & 0 \\[2ex]
0 & \dfrac{(Y^{(N)}_{12})^2\,\varphi^2}{2} & 0 \\[2ex]
0 & 0 & \dfrac{(Y^{(N)}_{33})^2\,\varphi^4}{16\,M_*^2}
\end{pmatrix}.
\end{equation}
\label{eq10}
Similarly, for the vector like quarks, we have
\begin{equation}
M_Q^2 =
\begin{pmatrix}
\dfrac{(Y^{(Q)}_{11})^2\,s^{2n}}{2^n (n!)^2 M_*^{2n-2}}
& 0 & 0 & 0 \\[2ex]
0 &
\dfrac{(Y^{(Q)}_{22})^2\,s^{2n}}{2^n (n!)^2 M_*^{2n-2}}
& 0 & 0 \\[2ex]
0 & 0 &
\dfrac{(Y^{(Q)}_{33})^2\,s^{2n}}{2^n (n!)^2 M_*^{2n-2}}
& 0 \\[2ex]
0 & 0 & 0 &
\dfrac{(Y^{(Q)}_{44})^2\,\varphi^2}{2}
\end{pmatrix}.
\label{eq11}
\end{equation}
Further, the breaking of the \(U(1)_a\) gauge symmetry will give rise to a massive gauge boson, which will have the field dependent mass as
\begin{equation}
    \label{eq12}
m_X^2
=
g_X^2\left[
q_S^2\left(s^2\right)
+
q_\phi^2\left(\varphi^2\right)
\right].
\end{equation}
where \(q_S=\frac{2}{n}\) and \(q_\phi = 6\) in our model.
Finally, the debye mass matrix for the scalars is given as
\begin{equation}
\Pi_{ij}(T)=
\frac{T^2}{24}\,
\mathrm{diag}\!\left(
\begin{aligned}
&216g_X^2+2(Y^{(N)}_{12})^2+6(Y^\mathrm{(Q)}_{44})^2+8\lambda_\phi+2\lambda_{\phi S},\\
&\frac{3}{2}g_X^2+8\lambda_S+2\lambda_{\phi S},\\
&216g_X^2+2(Y^{(N)}_{12})^2+6(Y^{(Q)}_{44})^2+8\lambda_\phi+2\lambda_{\phi S},\\
&\frac{3}{2}g_X^2+8\lambda_S+2\lambda_{\phi S}
\end{aligned}
\right),
\label{eq16}
\end{equation}
The appropriate charges for the various fields have been substituted already here, and \(n=4\). Since the Daisy term has an overall summation outside of it, we can take the trace of~\eqref{eq9} and~\eqref{eq16} to compute our Daisy corrected masses of the scalars, \[
\sum_i m_i^3(\phi,S,T) - m_i^3(\phi,S) \equiv \mathrm{Tr}\left[ \left(\mathcal{M}^\dagger\mathcal{M} + \Pi\right)^{3/2} \right] - \mathrm{Tr}\left[ \left(\mathcal{M}^\dagger\mathcal{M}\right)^{3/2} \right].
\] For our gauge boson, the Debye mass comes out to be
\begin{equation}
    \label{eq17}
\Pi_X(T)=g_X^2 T^2\left(\frac{100}{3}+\frac{4}{3n^2}\right).
\end{equation}

The benchmark points for this two field case are given below. We have taken all the Yukawa couplings to be of the same value, \(Y\). Further, we do not consider a case where the $\phi$ and $S$ VEVs have a large hierarchy between them.

\begin{figure}[H]
    \centering
    \includegraphics[width=0.7\linewidth]{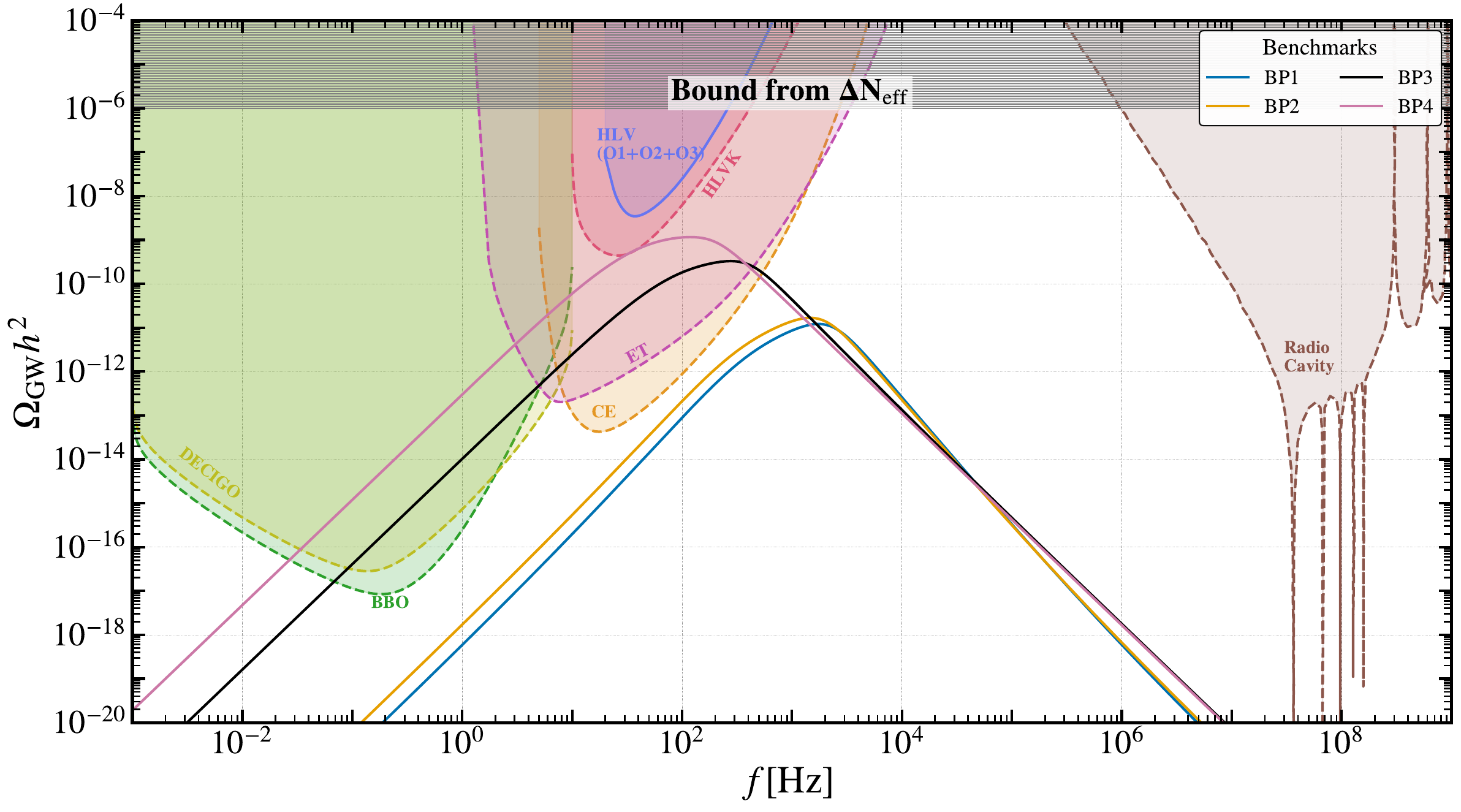}
    \caption{GW spectra in the non-conformal model for the benchmark points listed in Table \ref{tab:appen}.}
    \label{fig:GWNonConformalSpec}
\end{figure}

\begin{figure}[H]
    \centering
    \includegraphics[width=0.48\linewidth]{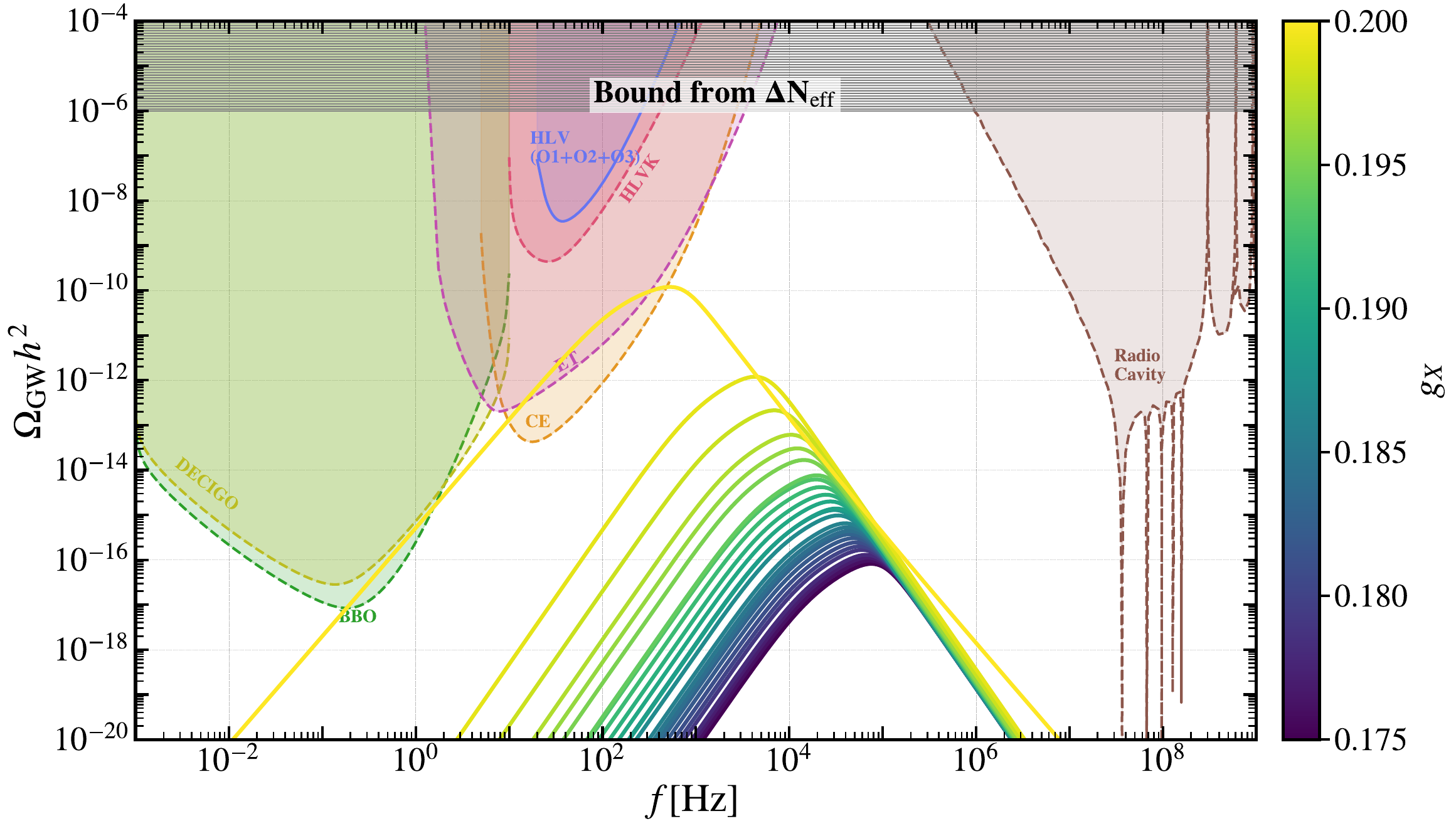}
    \includegraphics[width=0.48\linewidth]{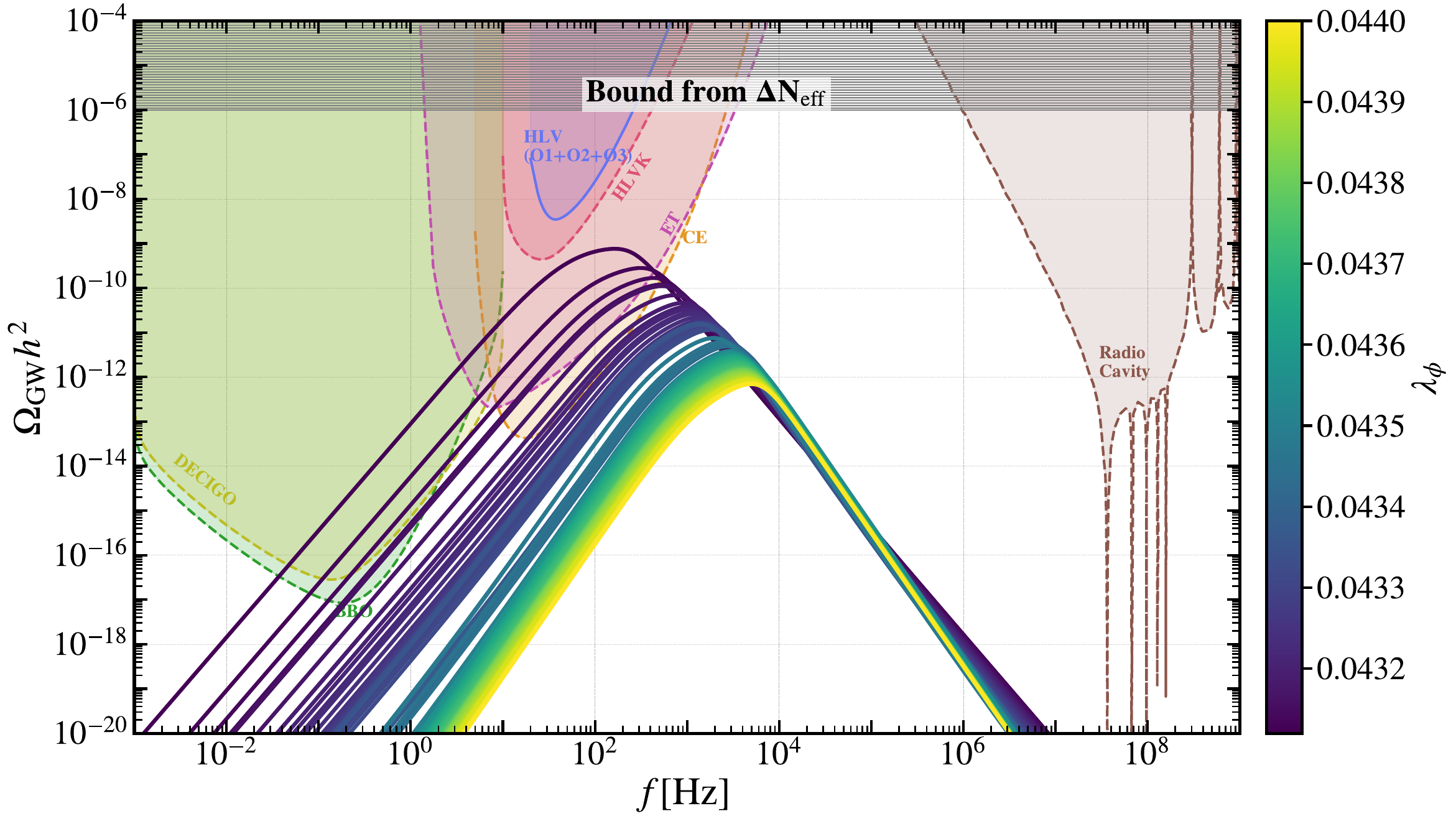}
    \caption{Variation in Gravitational wave spectra with change in model parameters. We have kept to the LO approximation for the determination of efficiency factors here.}
    \label{fig:GwNonConformalSpecScan}
\end{figure}

Further, we need to constrain these scales such that the PQ breaking scale \(f_a\) does not go below \(4 \times 10^8\) GeV, which is the lower limit on \(f_a\).
 We have,
 \begin{equation}
     f_a = \frac{f_\phi f_S}{\sqrt{f_\phi^2m^2n^2+f_S^2}}
     \label{eq18}
 \end{equation}
 where, \(m\) is the number of VLQs - 1 and also can be seen in the charge of the $\phi$ scalar which is written as \(2m\), \(n\) is from the charge of the S scalar, \(2/n\). In our case, we fix \((m,n)=(3,4)\). For 
 \[f_a \geq f_{a,0}=4\times10^8\text{ GeV},\]
 we get,
 \begin{equation}
     \label{eq19}
     f_s \geq \frac{mn f_\phi f_{a,0}}{\sqrt{f_\phi^2-f^2_{a,0}}}
 \end{equation}
 which immediately sets the constraint,
 \begin{equation}
     \label{eq20}
     f_\phi > f_{a,0}
 \end{equation}
 Further, imposing \(f_\phi > f_S\), we get
 \begin{equation}
     \label{eq21}
     f_\phi \geq f_{a,0}\sqrt{m^2n^2+1}
 \end{equation}
 which gives the inequality,
 \begin{equation}
     f_{a,0}\sqrt{m^2n^2+1}\leq f_S\leq f_\phi
 \end{equation}
For our choice of \((m,n)\), we have chosen \(f_\phi = 8\times10^9\) GeV, \(f_S = 5\times10^9\) GeV, giving us \(f_a = 4.16\times10^8\) GeV. We identify a few benchmark points consistent with the FOPT requirements for the non-conformal version of the model and list them in Table \ref{tab:appen}. The corresponding GW spectra for these benchmark points are shown in Fig. \ref{fig:GWNonConformalSpec}. Clearly, due to the lack of supercooling, the peak amplitude decreases noticeably compared to the supercooled FOPT scenario discussed in the main text. Fig. \ref{fig:GwNonConformalSpecScan} shows the variation of GW spectra with change in gauge coupling and scalar quartic coupling. Clearly, the peak amplitude remains smaller than the supercooled FOPT scenario irrespective of the choice of these couplings.

\begin{table}
\centering
\renewcommand{\arraystretch}{1.3}
\begin{adjustbox}{max width=\linewidth}
\begin{tabular}{|c|c|c|c|c|}
\hline
 \textbf{Parameter} & \textbf{BP1} & \textbf{BP2} & \textbf{BP3} & \textbf{BP4} \\
\hline
 $g_X$ & $0.2$ & $0.2$ & $0.2$ & $0.2$  \\
 $\lambda_\phi$ & $0.03828$ & $0.03825$ & $0.0381$ & $0.03809$ \\
 $\lambda_S$ & $0.1$ & $0.1$ & $0.1$ & $0.1$ \\
 $\lambda_{\phi S}$ & $0.02$ & $0.02$ & $0.02$ & $0.02$ \\
 $Y^{(Q)}_{11}$ & $0.1$ & $0.1$ & $0.1$ & $0.1$\\
 $Y^{(Q)}_{22}$ & $0.1$ & $0.1$ & $0.1$ & $0.1$ \\
 $Y^{(Q)}_{33}$ & $0.1$ & $0.1$ & $0.1$ & $0.1$ \\
 $Y^{(Q)}_{44}$ & $0.1$ & $0.1$ & $0.1$ & $0.1$ \\
 $Y^{(N)}_{12}$ & $0.1$ & $0.1$ & $0.1$ & $0.1$ \\
 $Y^{(N)}_{33}$ & $0.1$ & $0.1$ & $0.1$ & $0.1$ \\
 $M_\star$ & $4.42\times 10^{11}$ GeV & $4.42\times 10^{11}$ GeV & $4.42\times 10^{11}$ GeV & $4.42\times 10^{11}$ GeV \\
 $f_\phi$ & $8\times 10^{9}$ GeV & $8\times 10^{9}$ GeV & $8\times 10^{9}$ GeV & $8\times 10^{9}$ GeV \\
 $f_S$ & $5\times 10^{9}$ GeV & $5\times 10^{9}$ GeV & $5\times 10^{9}$ GeV & $5\times 10^{9}$ GeV \\
 $T_c$ & $3.39 \times 10^9$ GeV & $3.39\times10^9$ GeV & $3.38\times10^9$ GeV & $3.38\times10^9$ GeV \\
 $T_n$ & $1.18\times10^9$ GeV & $1.16\times10^9$ GeV & $9.95\times10^8$ GeV & $9.68\times10^8$ GeV \\
 $\alpha$ & 0.207 & $0.221$ & $0.389$ & $0.429$ \\
 $(\beta/H)_{S_3}$ & 38.083 & $32.639$ & $7.78$ & $3.87$ \\
\hline
\end{tabular}
\end{adjustbox}
\caption{Benchmark points consistent with FOPT in the non-conformal version of the quality axion model.}
\label{tab:appen}
\end{table}


\providecommand{\href}[2]{#2}\begingroup\raggedright\endgroup

\end{document}